\documentclass{ws-ijmpa}
\pdfoutput=1

\usepackage{amsmath,amssymb,mathtools,booktabs}
\usepackage{enumitem}
\usepackage{xcolor}
\usepackage{tikz}
\usetikzlibrary{arrows.meta,positioning,calc,decorations.pathmorphing,shapes.geometric}
\usepackage{hyperref}
\hypersetup{colorlinks=true,linkcolor=blue,urlcolor=blue,citecolor=blue}
\setlist[itemize]{leftmargin=2em}
\setlist[enumerate]{leftmargin=2em}

\makeatletter
\renewcommand\paragraph{\@startsection{paragraph}{4}{\z@}%
  {12\p@ \@plus 3\p@ \@minus 2\p@}{-.5em}{\paragraphfont}}
\makeatother
\newcommand{\Tr}{{\rm Tr}}
\newcommand{\tr}{{\rm tr}}
\newcommand{\Sym}{\operatorname{Sym}}
\newcommand{\sgn}{\operatorname{sgn}}

\newcommand{\dd}{\,{\mathrm d}}
\newcommand{\Adj}{{\mathrm{Adj}}}

\begin{document}
\markboth{R. de Mello Koch, M. Kim, A. L. Mahu and A. Rudra}{A Pedagogical Introduction to Invariant Theory and Finite-$N$ Holography}

\catchline{}{}{}{}{}
%
%%%%%%%%%%%%%%%%%%%%%%%%%%%%%%%%%%%%%%%%%%%%%%%%%%%%%%%%%%%%%%%%%%%%

\title{A pedagogical introduction to invariant theory and finite-$N$ holography}

\author{Robert de Mello Koch}

\address{School of Science, Huzhou Normal University, Huzhou 313000, China\\
\vspace{1mm}
\&
\vspace{1mm}
Mandelstam Institute for Theoretical Physics, School of Physics,\\
University of the Witwatersrand, Private Bag 3, Wits 2050, South Africa\\
robert@zjhu.edu.cn}

\author{Minkyoo Kim}

\address{School of Science, Huzhou Normal University, Huzhou 313000, China\\
\vspace{1mm}
\&
\vspace{1mm}
Research Institute of Basic Sciences, Seoul National University, Seoul 08826, Korea\\
mimkim80@gmail.com}

\author{Augustine Larweh Mahu}

\address{Department of Mathematics, University of Ghana, LG 62, Legon, Accra, Ghana\\
almahu@ug.edu.gh}

\author{Anik Rudra}

\address{School of Science, Huzhou Normal University, Huzhou 313000, China\\
\vspace{1mm}
\&
\vspace{1mm}
Mandelstam Institute for Theoretical Physics, School of Physics,\\
University of the Witwatersrand, Private Bag 3, Wits 2050, South Africa\\
anikrudra23@gmail.com}

\maketitle

\begin{abstract}
These lectures provide a pedagogical introduction to the study of finite-$N$ physics of the AdS/CFT correspondence. More specifically we develop the consequences of the trace relations for the space of gauge invariant operators, using powerful methods from invariant theory. Our key conclusions are 
\begin{itemize}
\item that the complete set of trace relations can be solved, leaving a non-redundant set of gauge invariant operators, 
\item that this non-redundant set can be generated using two types of generators, called primary and secondary invariants, and 
\item that the primary invariants acquire a natural interpretation as perturbative gravitational degrees of freedom, while the secondary algebra encodes non-perturbative effects of the dual gravitational theory.
\end{itemize}
\end{abstract}

\section{Introduction}

According to the AdS/CFT correspondence~\cite{AdSCFT}, theories of quantum gravity in asymptotically anti-de Sitter spacetimes are equivalent to ordinary quantum field theories in one lower dimension. The quantum field theories that arise in this correspondence are conformal field theories, and in some of the best understood examples, they are gauge theories. The strength of bulk gravitational interactions is controlled by $1/N^2$, where $N$ is the rank of the gauge group. Thus, the semiclassical limit of gravity is expected to emerge from the large-$N$ limit of the conformal field theory.

A particularly interesting dynamical regime is the finite-$N$ regime. At finite $N$, gravitational interactions are no longer parametrically weak, and one expects genuine quantum corrections to the classical spacetime geometry. The finite-$N$ theory also knows about intrinsically stringy and non-perturbative effects. For example, the spectrum of bulk fields is modified by the stringy exclusion principle~\cite{SEP,GGE}, and there is increasing evidence that black hole microstates are most naturally understood within the exact finite-$N$ Hilbert space~\cite{fort,evid}. On the conformal field theory side, this finite-$N$ physics is encoded in the trace relations: algebraic identities among gauge-invariant operators that are invisible in the strict large-$N$ limit. The goal of these lectures is to explore systematically the consequences of these trace relations for the structure of the theory, using the techniques of invariant theory~\cite{withantal,animik,minkyoo,anik,withjoao,withaniklarweh}.

In these lectures we focus on matrix quantum mechanics. These models provide simple but highly instructive laboratories in which the finite-$N$ trace relations can be analyzed explicitly. This allows us to extract sharp lessons about the structure of the Hilbert space, the organization of gauge-invariant states, and the non-perturbative content of the theory. Moreover, matrix quantum mechanics is not merely a toy model: such systems arise as important limits and subsectors of higher-dimensional conformal field theories\cite{SMT}. For this reason, the lessons obtained from these models are directly relevant to the finite-$N$ structure of the AdS/CFT correspondence.

\section{Loop Space at Finite $N$}

Three general references that we found helpful for this Section are Refs.\refcite{sturmfels}.

\subsection{Warmup: The algebra of gauge invariant operators for a single matrix}

Consider the matrix quantum mechanics of a single $N\times N$ matrix $X$, and assume that $X$ is Hermitian, $X=X^\dagger$. To keep the discussion general, for now we will leave the potential unspecified. The only thing we will require is that the model enjoys a $U(N)$ gauge symmetry under which the matrix $X$ transforms as
\begin{equation}
        X \longmapsto U X U^\dagger,\qquad U\in U(N) .
\end{equation}
The basic question we would like to answer is:
\begin{quote}
\emph{What is the complete space of gauge invariant operators?} 
\end{quote}
Under a unitary transformation, it is the eigenvalues of $X$ that are invariant. The space of all gauge invariants is the space of arbitrary polynomials of the eigenvalues $\lambda_i$ of $X$ and there are a total of $N$ eigenvalues. Equivalently, it is the space of arbitrary polynomials of the single traces
\begin{equation}
        m_n=\tr(X^n)=\sum_{i=1}^N(\lambda_i)^n,\qquad n=1,2,\ldots,N .
\end{equation}
Given the $N$ values $m_n$ we can solve for the eigenvalues $\lambda_i$, so this answer is equivalent to our first answer. This is called the singlet sector of the matrix model.
\begin{quote}
The complete space of gauge invariant operators is given by arbitrary polynomials of the \emph{generators}
$m_n$. 
\end{quote}
We say that the space of gauge invariant operators is generated by the single traces $m_n$ and that the generators $m_n$ are \emph{free}, meaning there are no relations between them. In the literature this space of gauge invariant operators is often called \emph{loop space}.

For example, if we set $N=2$ the complete set of generators is given by $\{{\rm Tr}(X),{\rm Tr}(X^2)\}$. One might be a little concerned because something like ${\rm Tr}(X^3)$ is a single trace and it is not among our generating invariants. It turns out that the ${\rm Tr}(X^3)$ can be written in terms of our set of generating invariants, thanks to the following \emph{trace relation}
\begin{equation}
2{\rm Tr}(X^3)-3{\rm Tr}(X^2){\rm Tr}(X)+{\rm Tr}(X)^3=0
\end{equation}
This trace relation is obeyed by \emph{any} $2\times 2$ matrix.

\subsection{The algebra of gauge invariant operators for $d$ matrices}

The single matrix problem is too simple to have much interesting structure. The multi-matrix problem is a lot richer and this is the problem that will keep us busy for most of these lectures. The question is: How do we describe the loop space of $d$ $N\times N$ matrices? We again have a Hermitian matrix model, and we again assume the model has a $U(N)$ gauge symmetry. The gauge symmetry acts on the matrices as follows  
\begin{equation}
        X^a\longmapsto U X^a U^\dagger,        \qquad a=1,2,\ldots,d .
\end{equation}
Again, for most of our discussion we will leave the potential arbitrary.

For the multi-matrix model invariants are given by taking traces of products of matrices
\begin{equation}
O={\rm Tr}(W)\qquad W=X^{a_1}X^{a_2}\cdots X^{a_k}
\end{equation}
We call the product of matrices $W$ a word. Cyclically related words give the same gauge invariant operator, thanks to cyclicity of the trace. The set of all cyclically distinct single trace operators generates the complete space of gauge invariant operators. At $N=\infty$ these generators are free -- there are no relations between them and the set
\begin{equation}
\{{\rm Tr}(X^a),{\rm Tr}(X^aX^b),{\rm Tr}(X^aX^bX^c),\cdots\}\label{setofgenerators}
\end{equation}
is a complete and non-redundant list of all gauge invariant operators. The set lists every independent single trace operator and the complete loop space is generated by taking any polynomial in these single trace generators. There is no description in terms of eigenvalues, as we had for a single matrix, simply because our collection of matrices does not commute and so can't be simultaneously diagonalized. Knowing the eigenvalues of each matrix is not enough information to construct the complete space of gauge invariant operators.

At finite $N$ the description is much richer. There are trace relations -- an infinite number of them, starting as soon as we have $N+1$ matrices in a single trace. That means that the set of generators \eqref{setofgenerators} becomes redundant. There are an infinite number of single trace operators in the set \eqref{setofgenerators} and an infinite number of trace relations. One might try to use each independent trace relation to eliminate a member of the set \eqref{setofgenerators}, and in this way obtain a complete and not redundant set of generating invariants. Some natural questions immediately come to mind: can we actually do this in practice and is the answer ``nice'' in any sense, i.e. does it have a definite algebraic structure that is easy to describe? Answering and understanding this question will keep us busy for the rest of these lectures.

To solve this problem we will need to use methods from invariant theory in mathematics. A very powerful tool from invariant theory that we will make use of again and again is the Hilbert series. We will now give a crash course in Hilbert series, developing only those aspects of the theory that we will need.

\subsection{Counting problems}\label{CountingProblems}
\smallskip

There is a motivational counting problem that will be useful for our goals:

\paragraph{Question:} How many independent monomials of degree $d$ can be constructed using $x$ and $y$?

\medskip

Of course there are an infinite number of possible monomials, so to get an interesting question we need to ask a slightly more refined question. The number of $x$'s and $y$'s you multiply to construct your monomial is called the degree of the monomial. A more interesting question is to ask: How many monomials of a given degree can we construct? Consider Table \ref{tab:monomials-two-variables} below which gives a list of monomials for low degree.

\begin{table}[htbp]
\centering
\begin{tabular}{c|c|c}
Degree & Polynomials & Number \\ \hline
0 & $1$ & 1 \\
1 & $x,y$ & 2 \\
2 & $x^2,xy,y^2$ & 3 \\
3 & $\qquad x^3,x^2y,xy^2,y^3\qquad$ & 4 \\
$\vdots$ & $\vdots$ & $\vdots$
\end{tabular}
\caption{Monomials in two variables organized according to their degree.}
\label{tab:monomials-two-variables}
\end{table}

This is the counting problem we want to solve. The key idea to solve this problem is to consider expanding the following product
\begin{align}
        {1\over 1-x}{1\over 1-y}
        &= (1+x+x^2+\cdots)(1+y+y^2+\cdots)\nonumber\\
        &=1+x+y+x^2+xy+y^2+x^3+x^2y+xy^2+y^3+\cdots .
\end{align}
From the last line above it is clear that we generate each monomial we want to count once and only once. To count, we can now set $x=t=y$. Hence
\begin{equation}
        {1\over (1-t)^2}=1+2t+3t^2+4t^3+\cdots .\label{countexpand}
\end{equation}
Comparing \eqref{countexpand} with Table \ref{tab:monomials-two-variables}, it is quite clear that the coefficient of $t^n$ counts how many degree $n$ polynomials we can construct. Here we are dealing with a polynomial ring freely generated by two generators which are $x$ and $y$. The generators are free because there is no relation which relates $x$ and $y$. The function
\begin{equation}
        H(t)={1\over (1-t)^2}
\end{equation}
is a simple example of a Hilbert series, which is a standard tool in the theory of graded rings. Here we see that the Hilbert series is actually the following function
\begin{equation}
H(t)=\sum_{i\in \{{\rm monomials}\}} t^{{\rm degree\,\,of\,\,}i}
\end{equation}
Here $\{{\rm monomials}\}$ is the set of all linearly independent monomials, $i$ indexes elements of the set, each of which has a definite degree. You can see that the denominator of the Hilbert series is $(1-t)(1-t)$ -- there are two terms multiplied because we have two generators, and each generator is of degree 1. To argue for this rule, it is helpful to do a few more examples. For our next example, imagine we want to count how many degree $d$ polynomials can be constructed using the generators
\begin{equation}
        x_1,\quad x_2,\quad x_3^2,\quad x_4^3 .
\end{equation}
The Hilbert series is
\begin{equation}
        H(t)={1\over (1-t)^2(1-t^2)(1-t^3)}=\sum_{n=0}^\infty c_n t^n .\label{SecondHilbert}
\end{equation}
Here $c_n$ is the integer that counts how many degree $n$ polynomials there are.

\paragraph{Exercise 1} Compute the coefficients $c_0$, $c_1$, $c_2$ and $c_3$ by explicitly constructing monomials. Compare your answer to the values you get by expanding the Hilbert series. 

\bigskip

The form of \eqref{SecondHilbert} supports the general rule we alluded to above: the Hilbert series has four factors in the denominator and the ring is generated by four generators. Further, we can read the degree of the generators from the four factors.

The rule that we have stated is true for \emph{freely generated rings}. This means the generators do not obey any constraints. However, there are many counting problems in which constraints are needed. We can illustrate the idea with a familiar problem: let's count the number of spherical harmonics in $\mathbb R^d$. Equivalently, we need to count the number of polynomials in the variables $x_1,x_2,\ldots,x_d$ subject to the constraint
\begin{equation}
        x_1^2+x_2^2+\cdots+x_d^2=1 . \label{SHconstraint}
\end{equation}
The reason why we need to supply this constraint is that the sum of the squares of the coordinates is invariant under rotations and thus this has angular momentum zero. However, we have already counted angular momentum zero when we counted ``1'' as the monomial of zero degree. The above constraint makes sure we don't count zero angular momentum twice. Applying our general rule you might argue that we have $d$-generators of degree 1, so that the Hilbert series should take the form
\begin{equation}
        H(t)={1\over (1-t)^d} .
\end{equation}
This is an overcount because, for example, we counted both 1 and the polynomial in \eqref{SHconstraint}, as well as any polynomial times 1 or the same polynomial times \eqref{SHconstraint}. Since we have over counted, to get the correct count we must subtract something off. The correct answer is obtained by subtracting one constraint of degree 2
\begin{equation}
        H(t)={1-t^2\over (1-t)^d} .
\end{equation}
You should be able to convince yourself that this subtraction is correcting things in the correct way. It is also easy to evaluate the above answer for various values of $d$ and make sure it matches what we expect by counting spherical harmonics. In $d=2$ dimensions we are expanding in waves on a circle. At every non-zero degree we have a non-zero momentum, and this may be a left moving wave and a right moving wave. At zero momentum there is a single mode -- the constant mode. This counting is reproduced by expanding the Hilbert series
\begin{align}
 d=2:        \qquad H(t)&=1+2t+2t^2+2t^3+\cdots .
 \end{align}
In $d=3$ the degree $l$ polynomials correspond to the usual spherical harmonics $Y^m_l(\theta,\phi)$, with the magnetic quantum number $m$ taking a total of $2l+1$ values: $m=-l,-l+1,\cdots,l-1,l$. Again this is perfectly reproduced by expanding the Hilbert series
\begin{align} 
 d=3:        \qquad H(t)&=1+3t+5t^2+7t^3+\cdots=\cdots +(2n+1)t^n+\cdots .
\end{align}        
Finally in $d=4$ the symmetry group $SO(4)$ can be traded for $SU(2)\times SU(2)$. You will be familiar with this identification which is often used to analyse the representations of the Lorentz group $SO(1,3)$ after Wick rotating to $SO(4)$. Polynomials of degree $n$ correspond to the $(n/2,n/2)$ representation of $SU(2)\times SU(2)$, which is equivalently a traceless symmetric tensor with $n$ indices. The spin $n/2$ irreducible representation of $SU(2)$ has dimension $2(n/2)+1=n+1$. Once again this counting is reproduced by expanding the Hilbert series
\begin{align} 
 d=4:\qquad H(t)&=1+4t+9t^2+16t^3+\cdots=\cdots+\left(2{\ell\over 2}+1\right)^2t^\ell+\cdots .
\end{align}
This is convincing evidence that we have indeed counted the number of spherical harmonics correctly.

At this point we can discuss the general form of a Hilbert series for a ring constructed using $n$ variables. As we have mentioned above, the presence of constraints implies that our generators are not all independent and counting as if they are all independent implies that we over count. To correct for this we need to subtract terms from the numerator, a term of the form $t^k$ for each constraint of degree $k$. However, the constraints themselves may not all be independent so that this subtraction overcorrects and we need to add some terms back in. Thus, for every relation, of degree $l$, between the constraints, we need to add a term in the numerator of the form $t^l$. But there may be relations between the relations between the constraints in which case we have added too much back in and we need to subtract something off again. The language that goes along with this discussion is as follows:
\begin{align}
\hbox{constraints} &= \hbox{first syzygies},\nonumber\\
\hbox{relations between constraints} &= \hbox{second syzygies},\nonumber\\
\hbox{relations between relations} &= \hbox{third syzygies},\nonumber\\
&\vdots\nonumber\\
&= n^{\rm th}\hbox{ syzygies}
\end{align}
``Syzygy'' originates from the ancient Greek word ``syzygia'', which means union or pair or yoke. One nightmare you might have is that this sequence is never-ending and new relations continue to emerge with no end. Notice however that  we have stopped the above list at $n^{\rm th}$ syzygies. This is a consequence of Hilbert's syzygy theorem which states that for a ring constructed using $n$ variables, we get at most $n^{\rm th}$ syzygies. Thus, the general structure of the Hilbert series is given by
\begin{eqnarray}
        H(t)&=&\big(1-\hbox{constraints}+\text{relations between constraints}\cr\cr
        &&-\text{relations between relations}+\cdots\big)
     {1\over \hbox{generator factors}} .
\end{eqnarray}
All of the rings that we will consider have free generators so the details of the general case will not be needed. Free generators are not constrained by any syzygies. 

Now we turn to our last and most important counting problem: How many polynomials can be constructed using $\{x,y\}$ freely and using $\{1,w^2,v^3\}$ exactly once? That is, one must choose exactly one of $1,w^2,v^3$. Following the logic of our first counting problem, we can generate all polynomials exactly once by expanding the following factors
\begin{eqnarray}
\frac{1}{(1-x)(1-y)}+\frac{w^2}{(1-x)(1-y)}+\frac{v^3}{(1-x)(1-y)}&=&1\,(1+x+y+x^2+xy+y^2+\cdots)\nonumber\\
+w^2\,(1+x+y+x^2+xy+y^2+\cdots)+&v^3&\!(1+x+y+x^2+xy+y^2+\cdots)
\end{eqnarray}
To count we now set
\begin{equation}
        x=y=t,\qquad w=v=t .
\end{equation}
The Hilbert series becomes
\begin{equation}
        H(t)={1+t^2+t^3\over (1-t)^2}.\label{HSModule}
\end{equation}
Here the numerator is for $\{1,w^2,v^3\}$, while the denominator is for $\{x,y\}$. The Hilbert series admits a power series expansion in powers of $t$
\begin{equation}
        H(t)=1+2t+4t^2+7t^3+\cdots=\sum_n c_n t^n .
\end{equation}
At degree 3 we have the monomials
\begin{equation}
        x^3, \quad x^2y, \quad xy^2, \quad y^3, \quad w^2x, \quad w^2y, \quad v^3 .
\end{equation}
There are 7 monomials explaining the $7t^3$ term in the Hilbert series. In general we again have that $c_n$ counts how many degree $n$ polynomials we can construct.

\subsection{Rings, Modules and Invariant Theory}

Let's now connect the concepts of rings and modules (which might not be so familiar) to the concepts of fields and vector spaces (which are familiar). Readable background references include Refs.\refcite{sturmfels}. A \emph{field} is a number system in which one can add, subtract, multiply, and divide by every nonzero element. Examples are $\mathbb{Q}$, $\mathbb{R}$, $\mathbb{C}$. A \emph{ring} has addition, subtraction, and multiplication, but division is not generally possible. Examples are $\mathbb{Z}$, $\mathbb{C}[x]$ and $\mathbb{C}[x_1,\ldots,x_n]$. The notation $\mathbb{C}[x]$ means polynomials in the variable $x$ with complex coefficients and the notation $\mathbb{C}[x_1,\ldots,x_n]$ means polynomials in the variables $x_1,\cdots,x_n$ with complex coefficients. Thus, a ring is obtained by weakening the axioms of a field: every field is a ring, but not every ring is a field.

The same relationship holds between vector spaces and modules. A vector space $V$ consists of objects that can be added and multiplied by scalars from a field $F$:
\begin{equation}
a v\in V,\qquad a\in F, \quad v\in V.
\end{equation}
A module $M$ is defined in exactly the same way, except that the scalars belong to a ring $R$:
\begin{equation}
r m\in M, \qquad r\in R, \quad m\in M.
\end{equation}
Let the module $M$ over the ring $R$ be generated by elements $m_1,\cdots ,m_n$. A relation among the generators is an equation
\begin{equation}
r_1m_1+\cdots +r_n m_n=0,\qquad r_i\in R.
\end{equation}
The relation is trivial when
\begin{equation}
r_1=\cdots =r_n=0.
\end{equation}
If there are no nontrivial relations, then $m_1,\cdots,m_n$ are linearly independent over $R$ and we are dealing with a \emph{free module}. Since they also generate $M$, every element $m\in M$ can be written uniquely as
\begin{equation}
m=r_1m_1+\cdots+r_n m_n\qquad r_i\in R
\end{equation}
In general a module need not possess a basis. A free module does have a basis given by the generators $m_1,\cdots ,m_n$. A free module behaves much like a vector space, except that its coefficients come from a ring rather than a field.

The last counting problem of Section \ref{CountingProblems} dealt with a free module. The module was defined over the ring freely generated by $x,y$ and the module basis was given by $\{1,w^2,v^3\}$. This identification helps to demystify the structure we found. In an ordinary three-dimensional vector space you would write a vector as
\begin{equation}
\vec{v}=c_1\hat{i}+c_2\hat{j}+c_3\hat{k}
\end{equation}
The coefficients $c_1$, $c_2$ and $c_3$ belong to a number field ($\mathbb R$ or $\mathbb C$) and raising these coefficients to a power is a perfectly legitimate operation. However, you would not raise elements of your basis (i.e. $\hat{i}$, $\hat{j}$ and $\hat{k}$) to a power. Likewise, it is perfectly legitimate to raise the generators $x$ and $y$ of the ring to any power, but you would not raise elements of the basis of the module $1$, $w^2$ and $v^3$ to a power. With this module structure in mind it should appear more natural why $1$, $w^2$ and $v^3$ always appear linearly and it is the same reason that $\hat{i}$, $\hat{j}$ and $\hat{k}$ appear linearly in vectors. Notice further that the denominator of the Hilbert series \eqref{HSModule} encodes the generators of the ring while the numerator encodes the module basis.

We will see that our set of invariants has the structure of a free module defined over a ring that is freely generated. The Hilbert series we are interested in for our study of the space of gauge invariant operators is given by
\begin{equation}
H(t)=\sum_{i}t^{{\rm degree \,\,of\,\, invariant\,\,}i}\label{HSforIR}
\end{equation}
where the sum is over any linearly independent basis of the space of gauge invariant operators. Computing this Hilbert series we will recover a Hilbert series of the same structure as shown in \eqref{HSModule}, so that we will learn what the module basis is from the numerator of the Hilbert series and we will learn what the generators of the free ring are from the denominator of the Hilbert series. To develop efficient methods to compute this Hilbert series, we now explore a natural connection to matrix quantum mechanics.

\subsection{Motivating matrix quantum mechanics}

We want to study the matrix quantum mechanics of $d$ $N\times N$ matrices $X^a$, $a=1,2,\cdots,d$. The matrix $X^a$ has a conjugate momentum $\Pi^a$ and there is a commutation relation which determines the operators
\begin{eqnarray}
[\Pi^a_{ij}, X^b_{kl}]&=&-i\hbar\, \delta^{ab}\delta_{jk}\delta_{il}\qquad\qquad\qquad i,j,k,l\,\,=\,\,1,\cdots,N,\quad a,b=1,\cdots,d.
\end{eqnarray}
In all subsequent formulas we set $\hbar=1$. The dynamics is determined by a Hamiltonian. For the case of a matrix harmonic oscillator (for example) the Hamiltonian is given by
\begin{eqnarray}
H&=&\frac12\sum_{a=1}^d {\rm Tr}\left[\Pi^a\Pi^a+X^aX^a\right]
\end{eqnarray}
We solve for the spectrum of this Hamiltonian using creation and annihilation operators in the usual way.

\paragraph{Exercise 2:} Introduce the creation and annihilation operators defined by
\begin{eqnarray}
A^a{}_{ij}&=&\frac{1}{\sqrt{2}}\left(X^a_{ij}+i\Pi^a_{ij}\right),\qquad
(A^a)^\dagger{}_{ij}\,\,=\,\,\frac{1}{\sqrt{2}}\left(X^a_{ij}-i\Pi^a_{ij}\right).
\end{eqnarray}
Verify that these operators obey the commutation relations
\begin{eqnarray}
[A^a{}_{ij},(A^a)^\dagger{}_{kl}]&=&\delta_{jk}\delta_{il}
\end{eqnarray}
and that the Hamiltonian can be written in terms of $A^a$ and $A^{a\dagger}$ as
\begin{eqnarray}
H&=&{\rm Tr}\left[(A^a)^\dagger A^a\right]+\frac{dN^2}{2}
\end{eqnarray}
where the constant $E_{\rm gs}=\frac{dN^2}{2}$ is the ground state energy. We will subtract this ground state energy off, which means we will simply drop the last term above. 
\bigskip

The ground state of the matrix oscillator $|0\rangle$ is defined as usual: it is the state annihilated by all of the annihilation operators $A^a_{ij}$
\begin{eqnarray}
A^a_{ij}|0\rangle&=&0\qquad\qquad\qquad a=1,\cdots,d,\quad i,j=1,\cdots,N.
\end{eqnarray}
This Hamiltonian has a gauge symmetry which acts as 
\begin{eqnarray}
X^a&\to&U X^aU^\dagger \qquad{\rm and} \qquad \Pi^a\,\,\to\,\, U\Pi^aU^\dagger
\end{eqnarray} 
Only states and observables that are invariant under this gauge symmetry are physical.

\paragraph{Exercise 3:} Argue that the most general eigenstate is given by acting on the ground state with traces of products of the creation oscillators, and that the energy of the state (after subtracting the ground state energy from the Hamiltonian) is equal to the number of creation oscillators
\begin{eqnarray}
|E=n\rangle={\rm Tr} (A^{a_1\dagger}\cdots A^{a_k\dagger}){\rm Tr} (A^{a_{k+1}\dagger}\cdots A^{a_l\dagger})\cdots{\rm Tr} (A^{a_q\dagger}\cdots A^{a_n\dagger})|0\rangle
\end{eqnarray}
(Hint: The multi-trace structure is dictated by the fact that we want a gauge invariant state so that all matrix indices must be summed. We could have had a single trace or any number of single traces on the right hand side -- it is only important that all indices are contracted.)
\bigskip

\noindent
Above we have discussed the oscillator, but as usual, we can choose to study any potential
\begin{eqnarray}
H&=&\frac12 {\rm Tr}\left[\sum_{a=1}^d \Pi^a\Pi^a+V(X^a)\right]
\end{eqnarray}
where the potential $V(X^a)$ is an arbitrary polynomial. As will become clear, the results that we obtain below for the structure of the space of gauge invariant operators will be applicable to any potential.

\subsection{Partition Functions and Hilbert Series}

We are going to compute the partition function of multi-matrix quantum mechanics. We will study harmonic oscillators and remove the ground state energy. Thus the energy of any state will be an integer,
\begin{equation}
        E_i=n_i .
\end{equation}
The partition function is
\begin{equation}
   Z(\beta)=\sum_i e^{-\beta E_i}=\sum_i (e^{-\beta})^{n_i}=\sum_i x^{n_i}, \qquad x=e^{-\beta},\label{PartFunc}
\end{equation}
where we can understand the sum over $i$ as a sum over a complete basis of energy eigenstates. Thus $Z(\beta)$ is counting states of energy $n_i$ in Hilbert space:
\begin{equation}
        Z(\beta)=\sum_m c_m x^m .
\end{equation}
Here $c_m$ is the number of states of energy $m$. This is starting to resemble the Hilbert series, which was counting polynomials of a given degree. We can make the analogy perfect as follows: recall from Exercise 3 that an energy eigenstate of energy $n$ takes the form
\begin{eqnarray}
|E=n\rangle&=&f(A^{a\dagger})|0\rangle\cr\cr 
f(A^{a\dagger})&=&{\rm Tr} (A^{a_1\dagger}\cdots A^{a_k\dagger}){\rm Tr} (A^{a_{k+1}\dagger}\cdots A^{a_l\dagger})\cdots{\rm Tr} (A^{a_q\dagger}\cdots A^{a_n\dagger}).
\end{eqnarray}
It is clear that $f(A^{a\dagger})$ is a gauge invariant operator. In fact, every gauge invariant operator corresponds to some energy eigenstate and every energy eigenstate corresponds to some gauge invariant operator -- there is a one-to-one correspondence between gauge invariant operators and energy eigenstates. In addition, the energy of the state is given by the degree of the invariant. Thus, the sum over states in the partition function \eqref{PartFunc} can be understood as a sum over invariants. Further the Boltzmann weight $e^{-\beta E_i}=x^{n_i}$ is the same as weighting each term by degree. Comparing with \eqref{HSforIR} it is clear that the partition function of the matrix oscillator is computing the Hilbert series of the invariant ring.

We want to take the trace relations into account. What is the form of the trace relations when written in this matrix oscillator problem? Recall that for $N=2$ we have a trace relation of the form
\begin{equation}
2{\rm Tr}((A^{a\dagger})^3)-3{\rm Tr}((A^{a\dagger})^2){\rm Tr}(A^{a\dagger})+{\rm Tr}(A^{a\dagger})^3=0\,.
\end{equation}
This becomes a statement that three states are no longer linearly independent
\begin{equation}
2{\rm Tr}((A^{a\dagger})^3)|0\rangle-3{\rm Tr}((A^{a\dagger})^2){\rm Tr}(A^{a\dagger})|0\rangle+{\rm Tr}(A^{a\dagger})^3|0\rangle=0\,,
\end{equation}
i.e. null states start to appear in the energy eigenbasis. Writing the partition function as a trace
\begin{equation}
Z(\beta)={\rm Tr}(e^{-\beta H})=\sum_{n}e^{-\beta E_n}\langle E_n|E_n\rangle
\end{equation}
it is clear that the null states do not contribute i.e. the trace relations have already been taken into account by the matrix oscillator partition function. This means that the Hilbert series it produces is telling us about the algebra of invariants \emph{after the trace relations have been imposed} -- and this is exactly the object we are interested in.

As we will explicitly derive below, we can write down a very concrete formula for the partition function of the matrix oscillator. Using this formula, what is the answer we obtain for the partition function? The generic form of the partition functions we compute is given by
\begin{equation}
        Z(x)=\frac{1+\sum_i c_i x^i}{\prod_j (1-x^j)} .
\end{equation}
This Hilbert series is familiar -- it is exactly the form expected for a free module -- and as a result, we know exactly how to interpret it.
\begin{itemize}
\item Each factor in the denominator corresponds to a generator of the ring. This is called a primary invariant. When generating the space of gauge invariant operators, primary invariants act freely i.e. they can be raised to any power. We will see that there are a total of $1+(d-1)N^2$ primary invariants, for $d\ge 2$. There are $N$ primary invariants for $d=1$.
\item Each monomial in the numerator corresponds to an element of the module basis and is called a secondary invariant. To generate the space of gauge invariant operators, each such gauge invariant operator is linear in secondary operators. Denote the total number of secondary invariants by $K$.
\end{itemize}
A linearly independent basis for the complete set of gauge invariant operators takes the form
\begin{equation}
\left\{P_1^{n_1}P_2^{n_2}\cdots P_{1+(d-1)N^2}^{n_{1+(d-1)N^2}}S_1,
P_1^{n_1}P_2^{n_2}\cdots P_{1+(d-1)N^2}^{n_{1+(d-1)N^2}}S_2,\cdots
P_1^{n_1}P_2^{n_2}\cdots P_{1+(d-1)N^2}^{n_{1+(d-1)N^2}}S_K\right\}.\label{HironakaBasis}
\end{equation}
This set includes terms with all possible non-negative values of the integers $n_1,n_2,\cdots,n_{1+(d-1)N^2}$. Notice that each term contains a single secondary invariant. Any gauge invariant operator is a unique linear combination of elements of the above set.

Since the trace relations are the constraints that determine the structure of the above set \eqref{HironakaBasis}, it is useful to review the general theory of trace relations. We will do that in the next subsection. This is followed by a derivation of the Molien-Weyl formula, which is an analytic formula for the matrix oscillator partition function.

\subsection{Trace Relations}

We have already given an example of a trace relation: any $2\times 2$ matrix $M$ obeys the following equation 
\begin{equation}
        \tr(M)^3-3\tr(M^2)\tr(M)+2\tr(M^3)=0 .\label{ElemTR}
\end{equation}
As will become clear from the derivation we give below, it does not matter if $M$ is Hermitian or unitary, or even if it can be diagonalized. To get some insight into why such a relation should hold, let us imagine that $M$ can be diagonalized and that its eigenvalues are denoted $\lambda_1$ and $\lambda_2$. Given the trace of $M$ and the trace of $M^2$ you can solve the following pair of equations
\begin{equation}
        \tr(M)=\lambda_1+\lambda_2,   \qquad   \tr(M^2)=\lambda_1^2+\lambda_2^2,
\end{equation}
to determine $\lambda_1$ and $\lambda_2$. And that means we know the value of
\begin{equation}
        \tr(M^3)=\lambda_1^3+\lambda_2^3 .
\end{equation}
This is the content of equation \eqref{ElemTR}, which expresses $\tr(M^3)$ in terms of $\tr(M)$ and $\tr(M^2)$. The trace relations are clearly a very important part of our analysis because they allow us to express traces of higher powers of a matrix in terms of polynomials in the traces of lower powers. 

How is the complete set of trace relations derived? We will start by considering $N=2$ dimensional matrices. The generalization to larger $N$ will be simple and immediate.  Towards this end, let $A,B$ and $C$ be three $2\times 2$ matrices. Start from the product of diagonal matrix elements $A_{ii}B_{jj}C_{kk}$ with no indices summed, and generate a sum of six terms by anti-symmetrising the column indices
\begin{align}
& A_{ii}B_{jj}C_{kk}-A_{ij}B_{ji}C_{kk}-A_{ii}B_{jk}C_{kj}\nonumber\\
&\quad -A_{ik}B_{jj}C_{ki}+A_{ij}B_{jk}C_{ki}+A_{ik}B_{ji}C_{kj}=0 .
\end{align}
The sum is zero because anti-symmetrising three indices which each take only two values always gives zero. You can easily check this by plugging in any values (1 or 2) for each of the three indices. If we now sum over $i,j$ and $k$ we find
\begin{align}
&\tr(A)\tr(B)\tr(C)-\tr(AB)\tr(C)-\tr(AC)\tr(B)\nonumber\\
&\quad -\tr(A)\tr(BC)+\tr(ABC)+\tr(ACB)=0 .\label{univident}
\end{align}
Procesi~\cite{procesi} has proved that this is a \emph{universal identity}. For a $d$-matrix model, by plugging arbitrary words into $A$, $B$ and $C$ we can generate the complete set of trace relations. In particular, \eqref{ElemTR} follows by setting $A=B=C=M$. If we choose 
\begin{equation}
        A=M,        \qquad B=M,  \qquad C=M^{n-2}
\end{equation}
we obtain a recursion of the form
\begin{align}
\tr(M^n)= {1\over 2}\big[&2\tr(M^{n-1})\tr(M)+\tr(M^2)\tr(M^{n-2}) -\tr(M)^2\tr(M^{n-2})\big] .\label{AllTrces}
\end{align}
This proves that if we are given the two single trace $\tr (M)$ and $\tr (M^2)$ we can generate the complete set of single traces -- so the trace relations explain why all higher power traces can be expressed in terms of these two. The corresponding statement for more than one matrix is less trivial and deriving it is one of the goals of these lectures.

The number of trace relations grows extremely rapidly with degree. The fact that it is possible to write a universal identity \eqref{univident} tells us that this growth is driven by the ever increasing number of words $A$, $B$ and $C$ with increasing degree.

The derivation of the trace relations for $N\times N$ matrices is the natural generalization. Choose $N+1$ words $W_1,W_2,\cdots,W_{N+1}$. Consider the product of diagonal elements $(W_1)_{i_1i_1}(W_2)_{i_2i_2}\cdots$ $(W_{N+1})_{i_{N+1}i_{N+1}}$ and anti-symmetrize over the column indices. The result, which is a sum over $(N+1)!$ terms, vanishes because anti-symmetrizing $N+1$ indices which each take $N$ values always vanishes. Summing over the indices $i_1,i_2,\cdots,i_{N+1}$ then gives a universal identity. By plugging in suitable choices for the $N+1$ words we can generate the complete set of trace relations for $N\times N$ matrices. 

Let us end this discussion with two general comments: the first trace relations for $N\times N$ matrices appear at degree $N+1$, and an $(N-1)\times (N-1)$ matrix obviously obeys\footnote{You can always add an $N^{\rm th}$ row and column of zeros to expand an $N-1$ dimensional matrix into an $N$ dimensional matrix.} the trace relations for $N\times N$ matrices. Thus, the trace relations obeyed by $N\times N$ matrices are a subset of the trace relations obeyed by $(N-1)\times (N-1)$ dimensional matrices.

\subsection{Molien-Weyl formula}

In this section our goal is to derive an explicit formula for the partition function of the matrix oscillator. Helpful references for this section include Refs.\refcite{4MW}. The partition function takes the form
\begin{equation}
        Z(x)=\sum_i x^{E_i}\times \#\{\hbox{states of energy }E_i\} .
\end{equation}
Every energy eigenstate is obtained by applying some number of oscillators to the ground state $|0\rangle$. Suppose we apply
\begin{equation}
\left.
\begin{array}{c}
 n_1\quad A^{\dagger}\hbox{'s}\,,\\
 n_2\quad A^{\dagger}\hbox{'s}\,,\\
 \vdots\\
 n_d\quad A^{\dagger}\hbox{'s}
\end{array}
\right\}
\qquad E=n_1+n_2+\cdots+n_d\equiv n
\end{equation}
to the ground state. Then
\begin{equation}
        Z(x)=\sum_{n_1=0}^{\infty}\cdots\sum_{n_d=0}^{\infty} x^{n_1}x^{n_2}\cdots x^{n_d}\,\# \label{MOPF}
\end{equation}
where $\#$ is the number of singlets created by the $n_a$ operators $A^{\dagger a}_{ij}$. The bulk of the work in this computation is in computing the integer $\#$. The logic is: work out the highly reducible representation produced by the $n_a$ creation operators and work out how many times the singlet appears. Let us state this representation-theoretically. The relevant tensor has the schematic form
\begin{equation}
        (A^{a_1 \dagger})_{i_1j_1}\,(A^{a_2 \dagger})_{i_2j_2}\cdots(A^{a_n \dagger})_{i_nj_n}\,|0\rangle,
\end{equation}
so it is a tensor with $2n$ indices. Each $(A^{a\dagger})_{ij}$ is in the adjoint representation, and because the creation operators commute (they are bosons), the tensor with $2n$ indices is in
\begin{equation}
        \Sym\big(\Adj^{\otimes n_1}\big)
        \otimes \Sym\big(\Adj^{\otimes n_2}\big)\otimes\cdots
        \otimes \Sym\big(\Adj^{\otimes n_d}\big)
        \equiv R_{n_1,n_2,\ldots,n_d} .
\end{equation}
This representation is highly reducible i.e. it is the direct sum of many different irreducible representations. Let us recall the structure of characters of direct sums of representations. If
\begin{equation}
        R=\Gamma_1\oplus \Gamma_2\oplus \Gamma_3,
\end{equation}
then, in a basis in which the representation matrices are block diagonal, we have
\begin{equation}
        \Gamma_R(g)=\begin{pmatrix}
        \Gamma_1(g)&0&0\\
        0&\Gamma_2(g)&0\\
        0&0&\Gamma_3(g)
        \end{pmatrix},
        \qquad d_R=d_1+d_2+d_3 .
\end{equation}
Therefore the character is given by
\begin{equation}
        \chi_R(g)=\tr(\Gamma_R(g))=\chi_1(g)+\chi_2(g)+\chi_3(g).
\end{equation}
The representation $R_{n_1,n_2,\ldots,n_d}$ will have the character
\begin{equation}
\chi_{R_{n_1,n_2,\ldots,n_d}}(g)=n_S\chi_S(g)+n_{R_1}\chi_{R_1}(g)+\cdots
\end{equation}
where we have used $S$ to label the singlet representation.  The numbers $n_S,n_{R_1},...$ are positive integers that count how many times the representations $S,R_1,\cdots$ appear when $R_{n_1,n_2,\ldots,n_d}$ is written as a direct sum over irreducible representations. Our goal is to determine the value of the integer $n_S$ -- this is the number we denoted by $\#$ in formula \eqref{MOPF}. We do this using character orthogonality. Character orthogonality gives, for a finite group,
\begin{equation}
        {1\over |G|}\sum_{g\in G}\chi_{r_1}(g)\chi_{r_2}(g^{-1})=\delta_{r_1r_2}.
\end{equation}
Equivalently, for a compact group
\begin{equation}
        \int_G \dd g\, \chi_{r_1}(g)\chi_{r_2}(g^{-1})=\delta_{r_1r_2} .
\end{equation}
Thus, using character orthogonality, we find for example
\begin{eqnarray}
\int_G dg \chi_{R_{n_1,n_2,\ldots,n_d}}(g)\chi_{R_1}(g^{-1})&=&\int_G dg(n_S\chi_S(g)+n_{R_1}\chi_{R_1}(g)+\cdots)\chi_{R_1}(g^{-1})\nonumber\\
&=&n_{R_1}
\end{eqnarray}
To obtain the integer we want, $n_S$, we need the character of the singlet representation. In the singlet, or trivial, representation we have
\begin{equation}
        \Gamma_S(g)=1\quad \forall g\in G, \qquad\Rightarrow\qquad \chi_S(g)=1\quad \forall g\in G .
\end{equation}
Thus the multiplicity of the singlet is obtained by integrating the character
\begin{eqnarray}
n_S&=&\int_G dg \chi_{R_{n_1,n_2,\ldots,n_d}}(g)\chi_S(g^{-1})\nonumber\\
&=&\int_G dg \chi_{R_{n_1,n_2,\ldots,n_d}}(g)
\end{eqnarray}

Thus, we now need to determine the character $\chi_{R_{n_1,n_2,\ldots,n_d}}(g)$ of the representation constructed by acting with the matrix creation operators. To accomplish this we will need the characters of tensor products. Recall that if $R$ is given by a tensor product of matrices
\begin{equation}
        R=\Gamma_1\otimes \Gamma_2\otimes \Gamma_3,
\end{equation}
then there is a basis labelled by the product of the indices of each matrix in the tensor product. Thus $R$ is a $d_R\times d_R$ matrix where
\begin{equation}
        d_R=d_1d_2d_3.
\end{equation}
It is helpful to think of the row and column labels of $R$ as the composite index $(i_1,i_2,i_3)$. The matrix of the representation $R$ given by the tensor product of representations $1\otimes 2\otimes 3$ is given by
\begin{equation}
        \Gamma_R(g)_{(i_1i_2i_3),(j_1j_2j_3)}=\Gamma_1(g)_{i_1j_1}\Gamma_2(g)_{i_2j_2}\Gamma_3(g)_{i_3j_3} .
\end{equation}
By taking a trace on both sides of this equation, it follows that the character of the tensor product is the product of characters
\begin{equation}
        \chi_R(g)=\sum_{i_1,i_2,i_3}\,\Gamma_R(g)_{(i_1i_2i_3),(i_1i_2i_3)}=\chi_1(g)\chi_2(g)\chi_3(g).
\end{equation}
The matrix creation operators transform in the adjoint representation. The transformation law for the adjoint representation is
\begin{equation}
        X^a\longmapsto U X^a U^\dagger, \qquad X^a_{ij}=U_{ik}X^a_{k\ell}(U^\dagger)_{\ell j} .
\end{equation}
Thus the adjoint representation matrix is
\begin{equation}
        \Gamma^{\rm adjoint}_{ij,k\ell}(U)=U_{ik}(U^*)_{j\ell} .
\end{equation}
Taking the trace we obtain the following character
\begin{equation}
        \chi_{\rm adjoint}(U)=\Gamma_{ij,ij}^{\rm adjoint}(U)=U_{ii}(U^*)_{jj}=\sum_{ij}\lambda_i\lambda_j^*
        =\sum_{ij}{\lambda_i\over \lambda_j} ,
\end{equation}
where we used the fact that the eigenvalues of a unitary matrix lie on the unit circle in the complex plane, so that $\lambda^*=\lambda^{-1}$. Since we take a product of matrix creation operators, we must evaluate the character of the tensor product of copies of the adjoint representation. Further, since the creation operators are bosons we are actually taking the symmetric product of copies of the adjoint. It is therefore necessary to discuss how to evaluate characters of the symmetric product of a particular representation $R$. Imagine we have a variable $X_i$ transforming under the representation $R$:
\begin{equation}
        X_i\longmapsto X'_i=\Gamma_R(g)_{ij}X_j,  \qquad i,j=1,\ldots,d_R .
\end{equation}
A product of $n$ such variables transforms under $\Sym(R^{\otimes n})$:
\begin{equation}
        X_{i_1}\cdots X_{i_n}
        \longmapsto (\text{Sym})_{i_1\cdots i_n,j_1\cdots j_n}
        \Gamma_R(g)_{j_1k_1}\cdots \Gamma_R(g)_{j_nk_n}X_{k_1}\cdots X_{k_n} .
\end{equation}
Here $(\text{Sym})_{i_1\cdots i_n,j_1\cdots j_n}$ is the projector onto the symmetric piece, given by
\begin{equation}
        (\text{Sym})_{i_1\cdots i_n,j_1\cdots j_n}={1\over n!}\sum_{\sigma\in S_n}
        \delta_{i_1j_{\sigma(1)}}\cdots \delta_{i_nj_{\sigma(n)}} .
\end{equation}
To develop some intuition for this formula, note that for $n=2$ this reduces to
\begin{equation}
        {1\over 2}\big(\delta_{i_1j_1}\delta_{i_2j_2}+\delta_{i_1j_2}\delta_{i_2j_1}\big)
\end{equation}
and therefore, when acting on a tensor we simply obtain the symmetric piece of the tensor
\begin{equation}
        \Sym_{ij,kl}T_{kl}={1\over 2}(T_{ij}+T_{ji}) .
\end{equation}
We can obtain a useful formula for $(\text{Sym})_{i_1\cdots i_n,j_1\cdots j_n}$ by invoking Wick's theorem, which at the level of ordinary Gaussian integrals says\footnote{${\cal N}$ is a normalization fixed by requiring that ${\cal N}\int \prod_i \dd z_i\dd \bar z_i\,e^{-\sum_j \bar z_j z_j}=1$.}
\begin{equation}
        {\cal N}\int \dd z_i\dd \bar z_i\,e^{-\bar z_i z_i}\,\bar z_{i_1}\cdots \bar z_{i_n}z_{j_1}\cdots z_{j_n}
        =\sum_{\sigma\in S_n}\delta_{i_1j_{\sigma(1)}}\cdots\delta_{i_nj_{\sigma(n)}}=n!\,\Sym .
\end{equation}
Putting these results together gives
\begin{eqnarray}
  \chi_{\Sym(R^{\otimes n})}(g)&=&{\rm Tr}\left((\Sym\,\,\Gamma_R(g)\otimes\Gamma_R(g)\otimes\cdots\otimes\Gamma_R(g))\right)\nonumber\\
   &=&{1\over n!}{\cal N}\int \prod_i \dd z_i\dd\bar z_i\, e^{-\sum_j \bar z_j z_j}\,\big(\sum_{k,l}\bar z_k\Gamma_R(g)_{kl}z_l\big)^n .
\end{eqnarray}
Using this result, we can now perform the following sum over $n$,
\begin{align}
\sum_{n=0}^\infty \chi_{\Sym(R^{\otimes n})}(g)t^n&={\cal N}\int \prod_i\dd z_i\dd \bar z_i\,e^{-\sum_j\bar z_j z_j}
\sum_{n=0}^\infty {t^n\over n!}\big(\sum_{k,l}\bar z_k\Gamma_R(g)_{kl}z_l\big)^n\nonumber\\
&= {\cal N}\int \prod_i \dd z_i\dd \bar z_i\,\exp\left[-\sum_j\bar z_j z_j+t\sum_{j,k}\bar z_j\Gamma_R(g)_{jk}z_k\right]\nonumber\\
&={1\over \det(1-t\Gamma_R(g))}=\exp\left[-\tr\log(1-t\Gamma_R(g))\right]\nonumber\\
&=\exp\left[\sum_{m=1}^\infty {1\over m}t^m\tr\big(\Gamma_R(g)^m\big)\right]\nonumber\\
&=\exp\left[\sum_{m=1}^\infty {1\over m}t^m\chi_R(g^m)\right] ,\label{SummedSymReps}
\end{align}
where we have freely used
\begin{equation}
        \log\det(\cdot)=\tr\log(\cdot), \qquad{\rm and}\qquad  -\log(1-x)=\sum_m {x^m\over m} .
\end{equation}
We will make good use of \eqref{SummedSymReps} below. 

This discussion has assembled all the ingredients we need to derive a closed formula for the partition function \eqref{MOPF}. The representation $R_{n_1,\ldots,n_d}$ is reducible so it can be written as a direct sum of many irreducible representations. The singlet multiplicity is the number of times the singlet representation $S$ appears in this direct sum
\begin{equation}
        R_{n_1\ldots n_d}=S\oplus S\oplus\cdots\oplus S\oplus R_1\oplus \cdots .
\end{equation}
As explained above, we extract the number of singlets by integration:
\begin{equation}
        \#=\int_{U(N)}\dd U\,\chi_{R_{n_1\cdots n_d}}(g)\chi_S(g)
        =\int_{U(N)}\dd U\,\chi_{R_{n_1\cdots n_d}}(g) .
\end{equation}
The character of the product representation is
\begin{align}
        \chi_{R_{n_1\cdots n_d}}(g)
        &=\chi_{\Sym(\Adj^{\otimes n_1})}(g)\cdots
          \chi_{\Sym(\Adj^{\otimes n_d})}(g) .
\end{align}
The partition function is then obtained by summing over the occupation numbers as follows
\begin{align}
        Z(x)&=\sum_{n_1=0}^\infty\cdots\sum_{n_d=0}^\infty x^{n_1}x^{n_2}\cdots x^{n_d}\,\# \nonumber\\
        &=\int_{U(N)}\dd U\sum_{n_1=0}^\infty \chi_{\Sym(\Adj^{\otimes n_1})}(U)x^{n_1} \cdots
        \sum_{n_d=0}^\infty \chi_{\Sym(\Adj^{\otimes n_d})}(U)x^{n_d} \nonumber\\
        &=\int_{U(N)}\dd U\left(\exp\left[\sum_{m=1}^\infty {1\over m}x^m\chi_{\Adj}(U^m)\right]\right)^d
\end{align}
where we have used \eqref{SummedSymReps} to obtain the final equality. The matrix $U$ has eigenvalues $\lambda_i$. It also depends on the angles of the unitary transformation that diagonalizes $U$. Since
\begin{equation}
        \chi_{\Adj}(U^m)=\sum_{i,j=1}^N {\lambda_i^m\over \lambda_j^m} ,
\end{equation}
the character depends only on the eigenvalues. After integrating over the angles,
\begin{equation}
        Z(x)=\left({1\over (1-x)^d}\right)^N {1\over N!(2\pi i)^N}
        \oint \prod_{j=1}^N {\dd\lambda_j\over \lambda_j}\,\Delta\bar\Delta\,{1\over F(\lambda_r)}
\end{equation}
where
\begin{equation}
   F(\lambda_r)=\prod_{1\le k<r\le N} \left(1-x{\lambda_r\over\lambda_k}\right)^d\left(1-x{\lambda_k\over\lambda_r}\right)^d .
\end{equation}
The Vandermonde determinant can be written as
\begin{equation}
        \Delta=\prod_{k<r}(\lambda_r-\lambda_k)=\sum_{\sigma\in S_N}\sgn(\sigma)\lambda_{\sigma(1)}^0\lambda_{\sigma(2)}^1\cdots \lambda_{\sigma(N)}^{N-1} .
\end{equation}
Now, notice that $F(\lambda_r)$ is invariant under permutations of the eigenvalues. By choosing a suitable permutation we can make each term in the product $\Delta\bar{\Delta}$ identical so that they sum to produce an $N!$ 
\begin{equation}
        \Delta\bar\Delta\longrightarrow N!\,\lambda_1^0\lambda_2^1\cdots\lambda_N^{N-1}\,\bar{\Delta}
        =N!\prod_{k<r}\left(1-{\lambda_r\over\lambda_k}\right) .
\end{equation}
It is perfectly legitimate to perform these different permutations for each separate term because both the measure of the integral and  $F(\lambda_r)$ are invariant under permutations of the eigenvalues. Therefore the partition function simplifies to
\begin{equation}
        Z(x)=\left({1\over (1-x)^d}\right)^N{1\over (2\pi i)^N}
        \oint_{S^1}\prod_{j=1}^N {\dd\lambda_j\over \lambda_j}
        \prod_{1\le k<r\le N}\left(1-{\lambda_r\over\lambda_k}\right)
        {1\over \left(1-x{\lambda_r\over\lambda_k}\right)^d
        \left(1-x{\lambda_k\over\lambda_r}\right)^d} .
\end{equation}
The eigenvalues of a unitary matrix have magnitude 1, $|\lambda_i|=1$, and this is why the integration contours above are unit circles. To simplify this result further, change variables to
\begin{equation}
        \lambda_j=t_j\cdots t_N .
\end{equation}
The Jacobian associated to this change of variables is 
\begin{equation}
        J=\det\left({\partial\lambda_i\over \partial t_j}\right)=t_1^{N-1}t_2^{N-2}\cdots t_{N-1}^{1}t_N^0
        ={\prod_{j=1}^N\lambda_j\over \prod_{j=1}^N t_j} .
\end{equation}
After relabelling $(t_i\to t_{i-1},\ t_1\to t_N)$ and integrating over $t_N$, we find
\begin{equation}
Z(x)=\left({1\over (1-x)^d}\right)^N{1\over (2\pi i)^{N-1}}\oint {\dd t_1\over t_1}\cdots\oint {\dd t_{N-1}\over t_{N-1}}
    \prod_{1\le k<r\le N-1}{1-t_{kr}\over (1-xt_{kr})^d(1-xt_{kr}^{-1})^d}\label{MWF}
\end{equation}
where
\begin{equation}
        t_{kr}=t_k t_{k+1}\cdots t_r .
\end{equation}
This is the basic formula we use to compute $Z(x)$. 

The entire derivation could have been done in the more general case in which we add chemical potentials $\mu_a$ for each species of field $X^a$. After adding chemical potentials, we obtain
\begin{align}
Z(x_1,x_2,\ldots,x_d)&=\left(\prod_{a=1}^d {1\over 1-x_a}\right)^N {1\over (2\pi i)^{N-1}}
\oint {\dd t_1\over t_1}\cdots\oint {\dd t_{N-1}\over t_{N-1}}\nonumber\\
&\quad\times \prod_{1\le k<r\le N-1}(1-t_{kr})\prod_{a=1}^d {1\over (1-x_a t_{kr})(1-x_a t_{kr}^{-1})},
\label{gradedMWF}
\end{align}
where
\begin{equation}
        x_a=e^{-\beta-\mu_a} .
\end{equation}

With this formula in hand, we can now describe in complete detail the loop space for a number of examples of $N\times N$ matrices with $N$ finite. The logic for our argument is always the same:
\begin{itemize}
\item We compute the partition function of the multi-matrix oscillators. This gives the Hilbert series of the invariant ring. From the Hilbert series we can read off a set of degrees for the primary and secondary invariants.
\item Using the above list of degrees, it is a straightforward exercise (at least for small values of $N$) to construct a complete set of generating invariants.
\item Using the trace relations, we can prove that polynomials in these invariants span all of loop space.
\end{itemize}
Since the proof uses nothing more than the trace relations, the statements we obtain are completely general. This last point is important: we have used the partition function of the matrix oscillator to obtain the Hilbert series. The Hilbert series is a property of the space of gauge invariant operators and is equally applicable for any potential we choose to consider. 

\section{Examples}

In this section we start by applying the methods developed in the last section to the one matrix model at any $N$ in Section \ref{onematx}, and then to the two and three matrix models in Sections \ref{twomatx} and \ref{threeematx}, both at $N=2$. In Section \ref{invtthry} we then review some general results from invariant theory which explain these examples and prove that our conclusions are general. The general formula for the number of primary invariants is then derived in Section \ref{NumPrimInv}. As $N$ increases, generating invariants that were originally secondary invariants can transition into primary invariants as demonstrated in Section \ref{IdentityChanging}. In Section \ref{TempInversion} we describe a temperature inversion symmetry (see~\cite{TIS} for a discussion) for the matrix oscillator partition function and its implications for the Hilbert series. This section closes with a count of the number of secondary invariants in Section \ref{CountSecondaries}.

\subsection{One matrix model}\label{onematx}

Consider first the matrix quantum mechanics of a single Hermitian matrix $X$. Using the Molien-Weyl formula, \eqref{MWF}, the single matrix partition function is:
\begin{equation}
        Z(x)={1\over (1-x)(1-x^2)(1-x^3)\cdots(1-x^N)} .
\end{equation}
The numerator immediately tells us that there are no secondary invariants besides the trivial secondary $\eta_0=1$ and the denominator, which is a product of $N$ factors, tells us that there are $N$ primary invariants. Further, each factor in the denominator tells us the degree of the corresponding primary invariant and this leads to an obvious guess for the corresponding operator,
\begin{equation}
        (1-x^n)\quad\longleftrightarrow\quad\tr(X^n) .
\end{equation}
For $N=2$ we only have $\tr(X)$ and $\tr(X^2)$ which connects to our discussion above.

From the obvious generalization of \eqref{AllTrces} we already know that any single trace operator can be recovered as polynomials in this primary set. The most general gauge invariant operator is a polynomial in the single traces, so that the above primary set does indeed generate the complete set of gauge invariant operators. Further, since all of these operators are needed to solve for the complete set of $N$ eigenvalues, we know that they are independent. Thus, we have achieved our goal of writing down a complete set of non-redundant generating invariants.

Before moving on it is worth stating this as a definite result for the Hilbert space of the matrix oscillator. Denote the oscillator constructed from $X_{ij}$ and its conjugate momentum $\Pi_{ij}$ by $A^\dagger_{ij}$. If we introduce the oscillators
\begin{equation}
A^\dagger_k\equiv \tr \left( (A^\dagger)^k\right)
\end{equation}
then the Hilbert space of the single matrix oscillator is the Fock space with generic state labelled by $N$ occupation numbers
\begin{equation}
|n_1,n_2,\cdots,n_N\rangle \,\,=\,\, (A_1^\dagger)^{n_1}(A_2^\dagger)^{n_2}\cdots (A_N^\dagger)^{n_N}|0\rangle
\end{equation}
In this sense, the primary invariants define Fock space oscillators.

\subsection{Two matrix model, $N=2$}\label{twomatx}

The model we will study next has two Hermitian matrices, which we denote as $X=X^1$ and $Y=X^2$. We will use the graded formula \eqref{gradedMWF} to compute the graded partition function with
\begin{equation}
        x=e^{-\beta-\mu_X},  \qquad  y=e^{-\beta-\mu_Y} .
\end{equation}
The result is obtained after a single integration which easily follows by applying the residue theorem
\begin{align}
Z(x,y)&=\left({1\over (1-x)(1-y)}\right)^2{1\over 2\pi i}\oint \dd t_1{t_1(1-t_1)\over (1-xt_1)(1-yt_1)(t_1-x)(t_1-y)}
\end{align}
The result is
\begin{equation}
        Z(x,y)= {1\over (1-x)(1-y)(1-x^2)(1-xy)(1-y^2)} .
\end{equation}
For this model we again only have the trivial secondary invariant $\eta_0=1$. From the denominator, which has five factors, we learn that there are five primary invariants. Since we have computed the completely graded partition function we can immediately read off how many $X$ and $Y$ matrices appear in each invariant. It is therefore simple to obtain the following list of primary invariants:
\begin{eqnarray}
        p_1=\tr(X),&\qquad& p_2=\tr(Y),\nonumber\\
        p_3=\tr(X^2),\qquad &p_4=\tr(XY)&,\qquad p_5=\tr(Y^2).\label{primarylist}
\end{eqnarray}
Notice that these five operators are all gauge invariant operators with length $\le 2$.

For a single matrix the eigenvalues give the complete list of gauge invariant information. For two or more matrices we can't assume that all the matrices can be simultaneously diagonalized. Consequently the complete set of gauge invariant information must include more than just the eigenvalues of the matrices. This is evident from our analysis here. The invariants $p_1$ and $p_3$ fix the eigenvalues of $X$, and the invariants $p_2$ and $p_5$ fix the eigenvalues of $Y$. The extra invariant $p_4$ includes information that is not captured by the eigenvalues alone.

 We know that these invariants are independent, i.e. that they are not related by any trace relation. This follows from the fact that for $N=2$  the first trace relation relates operators of degree 3 and our complete set of invariants is composed of invariants of degree 1 and 2. What we want to argue now is that this set is complete, i.e. that every gauge invariant operator can be written as some polynomial of the primary invariants.

At length three there are four independent gauge invariant operators
\begin{equation}
        \tr(X^3),\quad \tr(X^2Y),\quad \tr(XY^2),\quad \tr(Y^3) .
\end{equation}
Recall that the trace relation for $2\times2$ matrices is
\begin{align}
&\tr(A)\tr(B)\tr(C)-\tr(AB)\tr(C)-\tr(AC)\tr(B)\nonumber\\
&\quad -\tr(A)\tr(BC)+\tr(ABC)+\tr(ACB)=0 .\label{TRN2A}
\end{align}
By setting $A=B=C=X$, the trace relation becomes
\begin{equation}
        \tr(X)^3-3\tr(X)\tr(X^2)+2\tr(X^3)=0,
\end{equation}
so that
\begin{equation}
        \tr(X^3)={1\over 2}\left(3\tr(X)\tr(X^2)-\tr(X)^3\right)={1\over 2}(3p_1p_3-p_1^3) .
\end{equation}
Similarly,
\begin{itemize}
\item choosing $A=B=C=Y$, we learn that
\begin{equation}
        \tr(Y^3)={1\over 2}(3p_2p_5-p_2^3) .
\end{equation}
\item choosing $A=X$, $B=C=Y$, we learn that
\begin{equation}
        \tr(XY^2)={1\over 2}(2p_2p_4+p_1p_5-p_1p_2^2) .
\end{equation}
\item choosing $A=B=X$, $C=Y$, we learn that
\begin{equation}
        \tr(X^2Y)={1\over 2}(2p_1p_4+p_2p_3-p_1^2p_2) .
\end{equation}
\end{itemize}
Thus all operators with length $\le 3$ can indeed be written in terms of the primary invariants.

At each degree in $X$ and $Y$ we found a unique single-trace operator of length 3 and this was matched by a unique trace relation of the same multi-degree. This is not so simple at higher lengths. Consider an operator built from four fields. Choose two of the matrices to be $X$'s and two to be $Y$'s. There are \emph{two distinct} single-trace operators
\begin{equation}
        \tr(XYXY)   \qquad\hbox{or}\qquad  \tr(X^2Y^2) .
\end{equation}
Fortunately, this is matched by \emph{two independent} trace relations. With
\begin{equation}
        A=Y^2, \qquad B=X,  \qquad C=X,
\end{equation}
we obtain the following result for $\tr(X^2Y^2)$
\begin{equation}
        \tr(X^2Y^2)={1\over 2}\big(p_3p_5+2p_1p_2p_4-p_1^2p_2^2\big) .
\end{equation}
Next by taking
\begin{equation}
        A=XY,  \qquad B=X, \qquad C=Y .
\end{equation}
we obtain the following formula
\begin{equation}
        \tr(XYXY)={1\over 2}\big(p_2^2p_3+2p_4^2-p_3p_5+p_1^2p_5-p_1^2p_2^2\big) .
\end{equation}
What we need to prove is that there are always sufficient independent trace relations, obtained by making a suitable choice for $A,B$ and $C$ in \eqref{TRN2A}, to determine all single trace words.

We will give a proof by induction, which shows that all single-trace operators are indeed determined. Assume all traces with length $\le k$ can be written in terms of the primary invariants. With this assumption we will now prove all traces with length $\le k+1$ can be written in terms of the primary invariants. Consider the gauge invariant operator
\begin{equation}
        \tr(X^{n_1}Y^{m_1}), \qquad n_1+m_1=k+1\ge 4 .\label{typeone}
\end{equation}
It is clear that one of $n_1,m_1$ must be $>1$. Assume that it is\footnote{In the case that $n_1=1$, choose $A=Y^{m_1-1}$, $B=Y$ and $C=X^{n_1}$ and use the obvious generalization of the argument.} $n_1$, and choose
\begin{equation}
        A=X^{n_1-1}, \qquad B=X, \qquad C=Y^{m_1} .
\end{equation}
The trace relation gives
\begin{align}
&\tr(X^{n_1-1})\tr(X)\tr(Y^{m_1})-\tr(X^{n_1})\tr(Y^{m_1})-\tr(X^{n_1-1}Y^{m_1})\tr(X)\nonumber\\
&\quad -\tr(XY^{m_1})\tr(X^{n_1-1})+2\tr(X^{n_1}Y^{m_1})=0 .\label{typeonedeteremined}
\end{align}
All of these terms have length $\le k$, except the final term. Therefore, by the induction hypothesis, they can be expressed in terms of the primary invariants. Thus $\tr(X^{n_1}Y^{m_1})$ can be expressed in terms of the primary invariants.

Now consider the most general operator
\begin{equation}
        \tr(X^{n_1}Y^{m_1}X^{n_2}Y^{m_2}\cdots X^{n_q}Y^{m_q}),\qquad\sum_{i=1}^q (n_i+m_i)=k+1\qquad n_jm_j>0\,\,\forall\,\, j .
\end{equation}
We will call this a word of type $q$, because there are $q$ alternating $X^a Y^b$ blocks. In this classification, the invariant in \eqref{typeone} is of type 1. Take
\begin{equation}
        A=X^{n_1}, \qquad B=Y^{m_1}, \qquad C=X^{n_2}Y^{m_2}\cdots X^{n_q}Y^{m_q} .
\end{equation}
The trace relation will relate the type $q$ word to lower type words. Indeed, plugging this choice for $A,B$ and $C$ into \eqref{TRN2A}, the trace relation has the schematic form
\begin{eqnarray}
&&\tr(X^{n_1})\tr(Y^{m_1})\tr(X^{n_2}Y^{m_2}\cdots X^{n_q}Y^{m_q})
-\tr(X^{n_1}Y^{m_1})\tr(X^{n_2}Y^{m_2}\cdots X^{n_q}Y^{m_q})\nonumber\\
&&\quad  -\tr(X^{n_1+n_2}Y^{m_2}\cdots X^{n_q}Y^{m_q})\tr(Y^{m_1})
-\tr(X^{n_2}Y^{m_2}\cdots X^{n_q}Y^{m_q+m_1})\tr(X^{n_1})\nonumber\\
&&\quad +\tr(X^{n_1+n_2}Y^{m_2}\cdots X^{n_q}Y^{m_q+m_1})
+\tr(X^{n_1}Y^{m_1}X^{n_2}Y^{m_2}\cdots X^{n_q}Y^{m_q})\,\,=\,\,0 .
\end{eqnarray}
The first four terms in this trace have $\le k$ fields in each trace and hence are determined as polynomials in the primary invariants by the induction hypothesis. The fifth term is of type $q-1$ and the sixth term is of type $q$. Thus, if we can determine the type $q-1$ single trace operator, the above relation determines the type $q$ single trace operator. Given that we have determined the type 1 single trace operator in \eqref{typeonedeteremined}, the result follows by induction on $q$.

So for two $2\times 2$ matrices $X,Y$ we have proved that the complete space of gauge invariant operators is generated by \eqref{primarylist} and that this set of operators is independent. Thus, we have again achieved our goal of writing down a complete set of non-redundant generating invariants.

It is again worth stating this as a definite result for the Hilbert space of the two matrix oscillator. Denote the oscillator constructed from $X_{ij}$ and its conjugate momentum $\Pi_{ij}^X$ by $(A_X^\dagger)_{ij}$ and the oscillator constructed from $Y_{ij}$ and its conjugate momentum $\Pi_{ij}^Y$ by $(A_Y^\dagger)_{ij}$. Introduce the oscillators
\begin{equation}
        A_1^\dagger =\tr(A_X^\dagger),\qquad A_2^\dagger =\tr(A_Y^\dagger),
\end{equation}
\begin{equation}
        A_3^\dagger=\tr\left((A_X^\dagger)^2\right),\qquad A_4^\dagger=\tr(A_X^\dagger A_Y^\dagger),\qquad A_5^\dagger=\tr\left((A_Y^\dagger)^2\right).
\end{equation}
The Hilbert space of this two matrix oscillator is the Fock space with generic state labelled by $5$ occupation numbers
\begin{equation}
|n_1,n_2,\cdots,n_5\rangle \,\,=\,\, (A_1^\dagger)^{n_1}(A_2^\dagger)^{n_2}\cdots (A_5^\dagger)^{n_5}|0\rangle
\end{equation}
In this sense, the primary invariants again define Fock space oscillators.

\subsection{Three matrix model, $N=2$}\label{threeematx}

Next we study a model with three Hermitian matrices, which we denote as $X=X^1$, $Y=X^2$ and $Z=X^3$. This model is interesting as it will provide the first example with a non-trivial secondary invariant. We again compute the graded partition function using the Molien--Weyl formula \eqref{gradedMWF}. Towards this end introduce
\begin{equation}
        x=e^{-\beta-\mu_X},\qquad  y=e^{-\beta-\mu_Y},\qquad z=e^{-\beta-\mu_Z}.
\end{equation}
The result for the partition function is
\begin{equation}
        Z(x,y,z)=\frac{1+xyz}{(1-x)(1-y)(1-z)(1-x^2)(1-y^2)(1-z^2)(1-xy)(1-xz)(1-yz)}.
\end{equation}
The denominator has nine factors, so we learn that there are nine primary invariants. Thanks to the multi-grading we can immediately identify the following primary invariants
\begin{align}
        p_1&=\tr(X),      & p_2&=\tr(Y),      & p_3&=\tr(Z), \\
        p_4&=\tr(X^2),    & p_5&=\tr(Y^2),    & p_6&=\tr(Z^2), \\
        p_7&=\tr(XY),     & p_8&=\tr(XZ),     & p_9&=\tr(YZ).
\end{align}
The numerator of the partition function has two terms implying that there are two secondary invariants. Again, thanks to the grading we identify
\begin{equation}
        \eta_0=1,\qquad \eta_1=\tr(XYZ).
\end{equation}
Recall that in terms of these invariants, any gauge invariant operator ${\cal O}$ can be written as
\begin{equation}
{\cal O}=\sum_{n_1,\cdots,n_9=0}^\infty \left(a_{n_1,\cdots,n_9}\eta_0+b_{n_1,\cdots,n_9}\eta_1\right)\prod_{i=1}^9 p_i^{n_i}
\end{equation}
where $a_{n_1,\cdots,n_9}$ and $b_{n_1,\cdots,n_9}$ are complex numbers. We will not use the trace relations to prove that this set of operators is independent and complete. Rather, we would like to build some intuition as to why the secondary invariants only appear linearly. Once again, the trace relations determine this structure. Using the trace relations, it is possible to prove that there is a quadratic relation of the form
\begin{equation}
        (\eta_1)^2+p_I(p_i)\eta_1+p_{II}(p_i)\eta_0=0
\end{equation}
where $p_I$ is a degree $3$ polynomial and $p_{II}$ is a degree $6$ polynomial. The explicit forms of these polynomials are given in Refs.\refcite{withantal}. We never consider higher powers of the secondary invariants because all such higher powers can be rewritten as expressions which are linear in the secondary invariants. Thus, if we are to obtain a non-redundant set, we must consider only expressions which are linear in the secondary invariants.

Finally, it is again worth stating this conclusion as a definite result for the Hilbert space of the three matrix oscillator. Denote the oscillator constructed from $X$ and its conjugate momentum by $A_X^\dagger$, the oscillator constructed from $Y$ and its conjugate momentum by $A_Y^\dagger$ and the oscillator constructed from $Z$ and its conjugate momentum by $A_Z^\dagger$. Introduce the oscillators
\begin{align}
        A_1^\dagger&=\tr(A_X^\dagger),      & A_2^\dagger&=\tr(A_Y^\dagger),      & A_3^\dagger&=\tr(A_Z^\dagger), \\
        A_4^\dagger&=\tr\left((A_X^\dagger)^2\right),    & A_5^\dagger&=\tr\left((A_Y^\dagger)^2\right),    & A_6^\dagger&=\tr\left((A_Z^\dagger)^2\right), \\
        A_7^\dagger&=\tr(A_X^\dagger A_Y^\dagger),     & A_8^\dagger&=\tr(A_X^\dagger A_Z^\dagger),     & A_9^\dagger&=\tr(A_Y^\dagger A_Z^\dagger).
\end{align}
and define the states
\begin{equation}
|\eta_0\rangle=|0\rangle\qquad |\eta_1\rangle=\tr (A_X^\dagger A_Y^\dagger A_Z^\dagger)|0\rangle\,.
\end{equation}
The Hilbert space of this three matrix oscillator is the direct sum of two Fock spaces, each with a generic state labelled by $9$ occupation numbers, built on the two states $|\eta_0\rangle$ and $|\eta_1\rangle$
\begin{eqnarray}
|0;n_1,n_2,\cdots,n_9\rangle &=& (A_1^\dagger)^{n_1}(A_2^\dagger)^{n_2}\cdots (A_9^\dagger)^{n_9}|\eta_0\rangle\,,\nonumber\\
|1;n_1,n_2,\cdots,n_9\rangle &=& (A_1^\dagger)^{n_1}(A_2^\dagger)^{n_2}\cdots (A_9^\dagger)^{n_9}|\eta_1\rangle\,.
\end{eqnarray}
In this sense, the primary invariants again define Fock space oscillators, while the secondary invariants define non-perturbative states or sectors.

\subsection{General Comments}\label{invtthry}

The results that we have obtained in the last three sections all follow from general theorems from a field of mathematics known as invariant theory. Specifically, the decomposition of the algebra of gauge-invariant operators in terms of primary and secondary invariants is known as the Hironaka decomposition. Further, mathematicians have proved that if a ring is Cohen--Macaulay, it admits the Hironaka decomposition.

What does it mean for a ring to be Cohen--Macaulay? There are two ways to ``measure'' the dimension of the ring.
It is natural to define the dimension of the ring to be the largest number of algebraically independent generators you can write down. This is known as the Krull dimension of the ring and it is equivalently the number of primary invariants. For a $d$-matrix model with $d\ge 2$ the Krull dimension is given by
\begin{equation}
1+(d-1)N^2
\end{equation}
You can check that for $d=2$ and $N=2$ this is 5 which matches the number of primary invariants we found in Section \ref{twomatx}, while for $d=3$ and $N=2$ this is 9 which matches the number of primary invariants we found in Section \ref{threeematx}. The second way to measure the dimension of a ring is known as the depth of the ring. The depth of a ring is the number of equations you can impose before you encounter a hidden dependency. For example, consider the free ring
\begin{equation}
        R={\mathbb R}[x,y].
\end{equation}
which is the ring of polynomials in $x,y$ over the real number ${\mathbb R}$. The equations
\begin{equation}
        x=0, \qquad y=0
\end{equation}
are independent equations. Indeed,
\begin{align}
        x\,f(x,y)=0 &\quad\Rightarrow\quad f(x,y)=0, \\
        y\,f(x,y)=0 &\quad\Rightarrow\quad f(x,y)=0.
\end{align}
Thus this ring has depth 2. In addition, the largest number of algebraically independent generators is given by 2 since every element can be constructed from $x$ and $y$. Thus, the Krull dimension of this ring is 2. For this ring we thus see that these a priori different ways of measuring the dimension of the ring give the same answer
\begin{equation}
        \text{Krull dimension}=2=\text{depth}.
\end{equation}
This is a property shared by all free rings. Free rings have nice properties because they are very simple -- they are the rings that have free generators, i.e. their generators are not constrained by any relations. We define Cohen--Macaulay rings to be those rings for which
\begin{equation}
        \text{Krull dimension} = \text{depth of the ring}.
\end{equation}
The Cohen-Macaulay rings are the natural generalization of free rings that share the nice properties that free rings share. One important property for us, is that they admit a Hironaka decomposition.

Before continuing further, let us give an example of a ring that is not Cohen--Macaulay. Consider the ring
\begin{equation}
R=\mathbb{R}[x,y]/(x^2,xy).\label{eq:non-CM-ring}
\end{equation}
This means that we consider polynomials in $x$ and $y$, subject to the relations
\begin{equation}
x^2=0,\qquad xy=0.\label{eq:non-CM-relations}
\end{equation}
The variable $y$ remains unconstrained, so arbitrary powers $1,\, y,\, y^2,\, y^3,\,\ldots$ are nonzero and independent. Thus, the ring contains one continuous algebraic direction, parametrized by $y$. The variable $x$, on the other hand, does not provide a second independent direction because it is nilpotent: its square vanishes. Consequently, the Krull dimension of the ring is $1$. We now determine the depth. Using the relations in \eqref{eq:non-CM-relations}, a general element of the maximal ideal can be written as
\begin{equation}
f(x,y)=yg(y)+c x,
\end{equation}
where $g(y)$ is a polynomial and $c\in\mathbb{R}$. Multiplying by $x$, we obtain
\begin{equation}
x f(x,y)= xy g(y)+c x^2=0.
\end{equation}
Since $x$ itself is nonzero in ${\cal R}$ it is impossible to impose even one equation so the depth is zero. The depth and the Krull dimension do not agree, and hence ${\cal R}$ is not Cohen--Macaulay.

There is a simple physical picture for this example. The variable $y$ describes an ordinary algebraic direction along which one can move freely. The variable $x$ behaves instead like an infinitesimal fluctuation localized at $y=0$: it is nonzero, but both $x^2$ and $xy$ vanish. This nilpotent direction does not increase the geometric dimension of the space, but it introduces a hidden algebraic dependency. It is precisely this mismatch between the geometric dimension and the number of regular algebraic directions that causes the ring to fail to be Cohen--Macaulay.
 The Hochster--Roberts theorem~\cite{HochesterRoberts} proves that, for any linearly reductive group, the algebra of invariants is Cohen--Macaulay. A linearly reductive group is one for which every finite-dimensional representation of the group over a field of characteristic zero is completely reducible. All compact Lie groups are linearly reductive. Therefore all of the matrix gauge theories we will study in physics will admit the Hironaka decomposition. This implies that the results we have obtained above generalize for any matrix quantum mechanics model that we usually consider in physics. The Hironaka decomposition guarantees that the Hilbert series can be written in the form
\begin{equation}
H(x)\,\,=\,\,\frac{\sum_i c_i x^i}{\prod_j (1-x^j)^{n_j}}\label{HForm}
\end{equation}
with the coefficients $c_i$ and the powers $n_j$ given by non-negative integers.

There is an important point that we should state clearly: the list of primary and secondary invariants is not unique. Indeed, these invariants are determined by taking the full space of gauge invariant operators, and solving the trace relations. There are many more gauge invariant operators than there are trace relations, so this is an under determined system of equations. In this situation it is always up to us to choose which variables we choose to keep as parameters and which variables we solve for in terms of these parameters. Although the number of primary invariants is fixed by the Krull dimension, even the \emph{number} of secondary invariants will depend on these choices. A related comment applies to how we read the degree spectrum of primary and secondary invariants off of the Hilbert series. The Hilbert series is a rational function -- it is the ratio of a numerator polynomial and a denominator polynomial. Although the ratio is unique, the denominator and numerator themselves are not -- we might cancel common factors -- which is usually visible because some of the coefficients $c_i$ in \eqref{HForm} become negative. We might also multiply by factors of 1 in the form
\begin{equation}
\frac{1-x^k}{1-x^k}
\end{equation}
which, for a correct choice of $k$, does not change the structure given in \eqref{HForm}, but it does change the spectrum of degrees of the primary and secondary invariants.

To actually construct a list of primary and secondary invariants, we have found it useful to compute the partition function with a finer grading than grading by degree: we have been grading by the species of matrix. Although this has worked in the examples we have considered so far, it fails in general. The Hironaka decomposition is only consistent in general with a grading by total degree. We can explicitly see what goes wrong in general by considering the example of a matrix model with $d=4$ species and $N=2$. Denote the four matrices as $X^1=W$, $X^2=X$, $X^3=Y$ and $X^4=Z$. Given that we can write every operator in species-graded form and the trace relations respect species grading, it seems that the Hironaka decomposition must respect the finer grading. This is not the case. A valid list of primary invariants is given by
\begin{eqnarray}
        &&p_1=\tr(W), \qquad p_2=\tr(X), \qquad p_3=\tr(Y),\nonumber\\
        &&p_4=\tr(Z),  \qquad p_5=\tr(W^2),\qquad       p_6=\tr(X^2), \nonumber\\
        &&p_7=\tr(Y^2),\qquad   p_8=\tr(Z^2),\qquad    p_9=\tr(WX), \nonumber\\
        &&p_{10}=\tr(WY), \qquad p_{11}=\tr(XZ),\qquad p_{12}=\tr(YZ), \nonumber\\
        &&p_{13}=\tr(WZ)+\tr(XY).
\end{eqnarray}
Notice that, because of the form of $p_{13}$, these primary invariants do not respect the species grading. In fact, a more sophisticated argument shows that there is no choice of primary invariants that respects the species grading.

Another completely general property of the Hironaka decomposition is that the secondary invariants are quadratically reducible. They always obey a product rule of the form
\begin{equation}
\eta_\alpha\eta_\beta = \sum_\gamma C^\gamma_{\alpha\beta}(p_i)\eta_\gamma\label{SecProdRule}
\end{equation}
where the structure constants of this multiplication rule, $C^\gamma_{\alpha\beta}$ are polynomials in the primary invariants $p_\alpha$. This product rule is a direct consequence of the trace relations and it is the reason why the secondary invariants only ever appear linearly in the expansion of gauge invariant operators. Consequently, the complete space of gauge invariant operators is given by linear combinations of the operators
\begin{equation}
        \left\{ \eta_0\prod_{i=1}^{1+(d-1)N^2} p_i^{n_i}, \quad \eta_1\prod_{i=1}^{1+(d-1)N^2} p_i^{n_i},
        \quad \ldots,\quad \eta_{K-1}\prod_{i=1}^{1+(d-1)N^2} p_i^{n_i} \right\}
\end{equation}
where $K$ is the total number of secondary invariants and any non-negative power of the primary invariants is allowed.

\subsection{The number of primary invariants}\label{NumPrimInv}

In this section we would like to find a formula for the number of primary invariants. We will argue that the number of primary invariants is given by the number of free parameters remaining after a complete gauge fixing of the gauge symmetry. Towards this end, imagine we have $d$ $N\times N$ Hermitian matrices $X^a$. This is a total of $dN^2$ variables. Only a subset of these variables are invariant under the gauge symmetry:
\begin{equation}
        X^a\longmapsto U X^a U^\dagger,  \qquad X^a=(X^a)^\dagger.
\end{equation}
We will eliminate as many of the $dN^2$ variables as possible using the gauge transformations. This is a complete ``gauge fixing''. We will see that the number of parameters that remain agrees exactly with the number of primary invariants.

First, by using the gauge symmetry we can make $X^1$ diagonal:
\begin{equation}
        X^1=\begin{pmatrix} \lambda_1 &&& \\ & \lambda_2 && \\ && \ddots & \\ &&& \lambda_N \end{pmatrix}.
\end{equation}
This removes $N^2-N$ components. However, this does not completely fix the gauge symmetry because transformations of the form
\begin{equation}
        U=\begin{pmatrix}  e^{i\phi_1} & 0 & & \\  0 & e^{i\phi_2} & & \\ & & \ddots & \\ & & 0 & e^{i\phi_N}
        \end{pmatrix} = e^{i\phi_i}\delta_{ij}
\end{equation}
leave $X^1$ invariant. However, they act non-trivially on $X^2$:
\begin{equation}
        (U X^2 U^\dagger)_{ij} =U_{ik}X^2_{k\ell}U^\dagger_{\ell j} =X^2_{ij} e^{i(\phi_i-\phi_j)}.
\end{equation}
Thus the phases, with $\phi_1=\cdots=\phi_N$, cannot be used to remove any degree of freedom. The remaining $N-1$ transformations can be used to make the $N-1$ off the diagonal elements given by $X^2_{i\,i+1}$ and $X^2_{i+1\, i}$ real. Thus, we have managed to eliminate $N-1$ imaginary elements. This completely fixes the gauge symmetry, up to a total phase which leaves all matrices invariant, and we are thus left with
\begin{equation}
        dN^2-(N^2-N)-(N-1)=1+(d-1)N^2
\end{equation}
parameters after a complete gauge fixing. This agrees perfectly with
\begin{equation}
        \text{Krull dimension}=1+(d-1)N^2=\#\text{ of primary invariants}.
\end{equation}

Here we have illustrated that the number of parameters after gauge fixing agrees with the Krull dimension for matrix models, where the dynamical degrees of freedom transform in the adjoint representation of the gauge group. However, the conclusion is more general. For example, for vector models for which the vectors transform in the fundamental representation of the gauge group~\cite{animik}, the same conclusion can be demonstrated~\cite{minkyoo}.

\subsection{Identity changing}\label{IdentityChanging}

There is an interesting property that the space of gauge invariant operators exhibits, that goes under the name of \emph{stability}. Stability manifests itself in the counting of the number of gauge invariant operators. For $N\times N$ matrices, of any dimension $N$, the number of gauge invariant operators of degree $N$ or less agrees with the number of gauge invariant operators of the $N=\infty$ theory. This count of operators is stable -- it does not change as we pass from finite $N$ to $N=\infty$.

There is a version of stability that is exhibited in the degree spectrum of primary and secondary invariants. If certain primary degrees appear at one value of $N$ they persist at higher values of $N$. This can be illustrated with some examples in which we focus on $d=2$ matrices. For $N=4$ the Hilbert series is given by
\begin{equation}
        Z(x)=\frac{P_4(x)}{(1-x)^2(1-x^2)^3(1-x^3)^4(1-x^4)^6(1-x^5)^2},\label{Nis4HS}
\end{equation}
where the polynomial in the numerator takes the form
\begin{equation}
        P_4(x)=1+2x^5+\cdots+x^{24}.
\end{equation}
The first thing worth noting is that the number of factors in the denominator correctly reproduces the Krull dimension
\begin{equation}
        2+3+4+6+2=17=1+(4)^2.
\end{equation}
Notice that there are two primary invariants of length $5$. Now consider the Hilbert series obtained for $N=7$
\begin{eqnarray}
        Z(x)&=&\frac{P_7(x)}{(1-x)^2(1-x^2)^3(1-x^3)^4(1-x^4)^6(1-x^5)^8}\nonumber\\
        &&\times  \frac{1}{(1-x^6)^{11}(1-x^7)^8(1-x^8)^5(1-x^{10})^2(1-x^{12})},\label{Nis7HS}
\end{eqnarray}
and the polynomial in the numerator takes the form
\begin{equation}
        P_7(x)=1+3x^6+\cdots+x^{180}.
\end{equation}
Notice that once again the number of factors in the denominator correctly reproduces the Krull dimension
\begin{equation}
        2+3+4+6+8+11+8+5+2+1=50=1+7^2.
\end{equation}
To see the stability exhibited by the partition function notice that there are some common factors in the  denominator of \eqref{Nis4HS} and in the denominator of \eqref{Nis7HS}. For each such factor the power to which it is raised has either stayed the same or increased, supporting the idea that if a primary invariant appears at one value of $N$ it continues to appear at higher values of $N$. Notice that also for a single matrix we found that the generating invariants are all primary invariants, and that the largest invariant had a length of $N$. The Hilbert series for $N=4$ indicates that there are two primaries of length $5$, while at $N=7$ we have primaries of length $8,10,12$. Further, there were two secondary invariants of length $5$ at $N=4$, but increasing $N$ to $N=7$ both have changed character and become primary.

Generally, as $N$ increases, secondary invariants can transition to become primary invariants. With a little thought it is easy to argue that this had to happen. At $N=\infty$ there are no trace relations for any finite degree operator simply because the first trace relation appears at degree $N+1$. Thus, in the large-$N$ limit, loop space is freely generated and the Hilbert series is given by
\begin{equation}
        Z(x)=\frac{1}{\prod_i(1-x^i)^{n_i}},
\end{equation}
where $n_i$ is the number of single-trace operators formed using $i$ matrices. There are no secondary invariants, so that all secondary invariants that were present at finite $N$ have transitioned to become primary invariants.

This identity changing property of the secondary invariants is reminiscent of fortuitous operators~\cite{fort}. There is mounting evidence that fortuitous operators in CFT are dual to microstates~\cite{evid} of a $\frac{1}{16}$-BPS black hole. Fortuitous operators in ${\cal N}=4$ super Yang-Mills theory, are $\frac{1}{16}$-BPS operators, whose saturation of the BPS bound relies on a trace relation. When $N$ is increased by a single unit, the trace relation no longer holds and the operator fails to be a black hole microstate. We will see further parallels between fortuitous states and secondary operators below. 

\subsection{Temperature inversion symmetry}\label{TempInversion}

Another interesting property of the Hilbert series for the multi-matrix model invariants is that its numerator is palindromic~\cite{paledrmic}. A palindrome is a word, phrase, number, or other sequence of characters that reads the same forward and backward. A good example of a palindrome is ``racecar''. A palindromic polynomial has coefficients that are the same whether you read from smallest to largest degree, or from largest to smallest 
\begin{equation}
        P(x)=a_0+a_1x+a_2x^2+\cdots+a_nx^n, \qquad a_k=a_{n-k}, \qquad k=0,1,2,\ldots,n.
\end{equation}
Consequently, for a palindromic polynomial, we have the property
\begin{equation}
\begin{aligned}
        P(x) &=x^n\left(a_nx^{-n}+a_{n-1}x^{1-n}+\cdots+a_1x^{-1}+a_0\right) =x^n P\left(\frac1x\right).\label{NTrans}
\end{aligned}
\end{equation}
As an example, the Hilbert series for the matrix model of $d=4$ $2\times 2$ matrices is given by
\begin{equation}
        H(x)=\frac{1+x^2+4x^3+x^4+x^6}{(1-x)^4(1-x^2)^9}.
\end{equation}
The numerator is clearly a palindromic polynomial. Also, we again see that the number of factors in the denominator agrees with the Krull dimension
\begin{equation}
        4+9=13=1+3\cdot(2)^2=1+(d-1)N^2.
\end{equation}
In the single matrix model, all single trace invariants with length $\le N$ are included as primary invariants. For this multi-matrix model at $N=2$ there is a secondary invariant of length $2$.

At the level of the Hilbert series there is a nice generalization of the property \eqref{NTrans}. Under $x\to x^{-1}$ the primary factors in the denominator have the following transformation property
\begin{equation}
        \frac{1}{1-x^i}= -\frac{1}{x^i}\left(\frac{1}{1-x^{-i}}\right).
\end{equation}
Since the partition function is a product of $1/(1-x^i)$ factors times a palindromic polynomial, $H(x)$ and $H(x^{-1})$ differ by a power of $x$ and possibly a sign. It turns out that, for $d=2$,
\begin{equation}
        Z(x)=x^{N^2}H(x)\label{RHS}
\end{equation}
obeys
\begin{equation}
        Z(x^{-1})=(-1)^{N-1}Z(x).\label{ZTRansf}
\end{equation}
This elegant transformation law only appears after we define $Z(x)$ as a monomial times the Hilbert series as in \eqref{RHS}. How should we interpret $x^{N^2}$? Since
\begin{equation}
        x^{N^2}=e^{-\beta N^2}=\left(e^{-\beta/2}\right)^{2N^2},
\end{equation}
this factor is nothing but the ground-state contribution ($=\frac12$ for each of the $2N^2$ oscillators) to the energy. Thus, \eqref{RHS} is the relation between the Hilbert series and the physical partition function with ground state energy included. What is the physical meaning of the ``symmetry'' exposed in \eqref{ZTRansf}? Recall that
\begin{equation}
        x=e^{-\beta},  \qquad \beta=\frac1T.
\end{equation}
Therefore
\begin{equation}
        x\longrightarrow \frac1x   \quad\Rightarrow\quad  T\longrightarrow -T.
\end{equation}
Thus,  \eqref{ZTRansf} tells us that the partition function has temperature inversion symmetry.

The fact that the Hilbert series is palindromic tells us something about the underlying ring: the ring is usually Gorenstein. A Gorenstein ring is a ring that, although it may have singularities, is sufficiently well behaved that it admits a version of Poincare duality: a symmetry between functions and holes. The analogy with Poincare duality implies palindromicity, a duality between the relevant cohomology group and the $n$-th homology group.

\subsection{How many secondary invariants are there?}\label{CountSecondaries}

We have already seen that the number of primaries corresponds to a nice property of the ring: it is the Krull dimension. Is the number of secondary invariants also related to a nice property of the ring? We know that the answer is no. The number of secondary invariants is not a property of the ring -- it changes depending on the choice made for the primary invariants. So before the homogeneous system of parameters is specified, the number of secondary invariants is not determined. We will make the natural choice that we choose the set of primaries so that each has the lowest possible degree. With this choice the number of secondary invariants is fixed and we can determine how this number grows with $N$. That is the goal of this section.

To get some insight into the number of secondary invariants, we can gather some data by computing the Hilbert series for $d=2$ matrices, with a range of values of $N$. The number of factors in the denominator must give the Krull dimension, which for $d=2$ is $1+N^2$. The numerator of the Hilbert series, evaluated at $x=1$, then gives the number of secondary invariants. The table below summarizes this data starting from $N=2$ and proceeding to $N=7$.

\bigskip

\begin{table}[h]
\centering
\renewcommand{\arraystretch}{1.2}
\setlength{\tabcolsep}{12pt}
\begin{tabular}{|c|r|r|}
\hline
$N$ &
\multicolumn{1}{c|}{Primary invariants} &
\multicolumn{1}{c|}{Secondary invariants} \\
\hline\hline
2 &  5 & 1 \\
\hline
3 & 10 & 2 \\
\hline
4 & 17 & 64 \\
\hline
5 & 26 & 15,624 \\
\hline
6 & 37 & 312,606,720 \\
\hline
7 & 50 & 21,739,438,196,736 \\
\hline
\end{tabular}
\caption{Numbers of primary and secondary invariants for several values of $N$ and $d=2$.}
\label{tab:number-of-invariants}
\end{table}

\bigskip

The growth in the number of primary invariants is a power, $N^2$. Comparing these numbers, it is clear that the growth of secondary invariants is much more rapid than a power. Although we don't have many data points to fit, the suggested numerical behaviour is
\begin{equation}
        N_{\text{secondary}}\approx e^{1.1N^2-3.2N}.
\end{equation}
i.e. the growth is as $e^{cN^2}$ with $c$ a number that is fixed as $N\to\infty$. 

We can give a careful proof that the number of secondary invariants does indeed grow this fast~\cite{minkyoo}. To do this, we return to the $d=2$ matrix oscillator. It is possible to write down a complete set of energy eigenvalues and eigenstates. Finding the energy eigenvalues is a simple task. It is harder to find the degeneracy of each level and harder still to select a basis for these energy eigenstates. We won't derive these eigenstates -- we will simply state what the result is and give a reference to the literature. We again subtract off the ground state energy. A complete basis for the energy eigenstates at energy level $n+m$ is given by the restricted Schur polynomials~\cite{RSP}
\begin{equation}
\chi_{R,(r,s)\alpha\beta}(A^{1\dagger},A^{2\dagger})
\end{equation}
The labels above are three Young diagrams: $R$ has $n+m$ boxes, $r$ has $n$ boxes and $s$ has $m$ boxes. We can think of $R$, $r$ and $s$ as labels for irreducible representations of $U(N)$. The labels $\alpha$ and $\beta$ are multiplicity labels and they run from over $1,2,\cdots,f_{rsR}$. Recall that the Littlewood-Richardson number is a non-negative integer that tells us how many times irreducible representation $R$ appears in the product of representations $r\times s$. For a pedagogical introduction to restricted Schur polynomials see~\cite{RSPintro}

Using this explicit answer for the energy eigen problem, we can determine a lower bound for the partition function. Then, by using the Hironaka form of the partition function, we provide an upper bound for the contribution from the primary invariants. The difference between the upper bound of the primary contribution and the lower bound of the partition function must be made up by the secondary invariants -- and this tells us how the number of secondary invariants grows as $N$ is increased.\footnote{Although we have used the restricted Schur polynomials in this computation, there are a number of other bases that could also have been used -- see~Refs.\refcite{Sanjaye}.}

We now make this argument precise.  Let
\begin{equation}
 Z(x)=\sum_{q\geq 0}c_q^{\rm T}x^q,
\end{equation}
where the superscript ${\rm T}$ emphasizes that $c_q^{\rm T}$ counts the total number of independent gauge invariant operators of degree $q$. The counting of states in the restricted Schur basis gives
\begin{equation}
 c_{n+m}^{\rm T}=\sum_{\substack{R\vdash n+m\\ \ell(R)\leq N}}\sum_{r\vdash n}\sum_{s\vdash m}
 \left(f_{rs}^{R}\right)^2.
\end{equation}
The square appears because both multiplicity labels $\alpha$ and $\beta$ run from $1$ to $f_{rs}^{R}$.  Thus, the Littlewood--Richardson coefficients directly measure the degeneracy of the energy eigenstates. We now focus on operators of degree $q=\alpha N^2$, where $\alpha$ is held fixed as $N\to\infty$.  The largest Littlewood--Richardson coefficient associated with Young diagrams containing a total of $q$ boxes has the asymptotic behaviour~\cite{PakPanova}
\begin{equation}
 f_{rs}^{R} = \exp\left(\frac{q}{2}\log 2+O(\sqrt{q})\right).
\end{equation}
Keeping only the single term in the restricted Schur sum that contains this largest coefficient gives a lower bound on the total number of operators,
\begin{equation}
 c_{\alpha N^2}^{\rm T} > \exp\left(\alpha\log 2\,N^2+O(N)\right).
\end{equation}
The Young diagrams which maximize the Littlewood--Richardson coefficient~\cite{PakPanova} approach the Vershik--Kerov--Logan--Shepp shape~\cite{VKLS}.  For the small fixed values of $\alpha$ that we use
below, these diagrams fit inside the finite-$N$ restriction that $R$ has at most $N$ rows.
Thus the finite-$N$ cutoff does not modify the leading exponential behaviour.

We know that the Hilbert series takes the Hironaka form. Using this form, it is useful to separate the Hilbert series into the contribution generated by the primary invariants and the contribution from the secondary invariants.  Write
\begin{equation}
 Z(x)=Z_{\rm p}(x)Z_{\rm s}(x), \qquad Z_{\rm p}(x)=\prod_{j\geq 1}\frac{1}{(1-x^j)^{c_j^{\rm p}}}=\sum_{q\geq 0}d_qx^q, \qquad  Z_{\rm s}(x)=1+\sum_{i\geq 1}c_i^{\rm s}x^i.
\end{equation}
Here $c_j^{\rm p}$ is the number of primary invariants of degree $j$, while $c_i^{\rm s}$ is the number of secondary invariants of degree $i$.  For the two-matrix model, we have
\begin{equation}
 \sum_{j\geq 1}c_j^{\rm p}=N^2+1, \qquad N_{\rm secondary}=1+\sum_{i\geq 1}c_i^{\rm s}.
\end{equation}
We will compare the total number of operators at degree of order $N^2$ with the number that can be generated using the primary invariants alone. To estimate how many operators can be generated using only the primary invariants, recall that
\begin{equation}
 Z_{\rm p}(x) = \prod_{j\geq 1}\frac{1}{(1-x^j)^{c_j^{\rm p}}}=\sum_{q\geq 0}d_qx^q.
\end{equation}
The coefficient $d_q$ counts the number of monomials of total degree $q$ that can be constructed from the primary invariants.  We do not know all of the integers $c_j^{\rm p}$, so it is difficult to compute $d_q$ exactly.  Fortunately, we only need an upper bound. To derive the required bound, choose a fixed positive integer $k$ and define
\begin{equation}
 C_k=\sum_{j=1}^{k-1}c_j^{\rm p}.
\end{equation}
There are $C_k$ primary invariants of degree less than $k$, while the remaining $N^2+1-C_k$ primary invariants have degree at least $k$.  We can overcount the monomials of a fixed degree by replacing every primary of degree less than $k$ by a primary of degree $1$, and replacing every remaining primary by a primary of degree $k$.  This gives the auxiliary partition function
\begin{equation}
\widehat Z_{\rm p}^{(k)}(x)=\frac{1}{(1-x)^{C_k}}\frac{1}{(1-x^k)^{N^2+1-C_k}}=\sum_{q\geq 0}\widehat d_q^{(k)}x^q,
\end{equation}
with $d_q\leq \widehat d_q^{(k)}$. The reason for this inequality is simple.  Lowering the degree assigned to a generator can only make it easier to construct monomials of a prescribed total degree.  Any degree lost by making these replacements can be supplied by one of the degree-one primary invariants. The coefficient in the auxiliary partition function is
\begin{equation}
 \widehat d_q^{(k)} = \sum_{b=0}^{\lfloor q/k\rfloor} \binom{q-kb+C_k-1}{C_k-1} \binom{b+N^2-C_k}{N^2-C_k}.
\end{equation}
For fixed $k$, the stability of the low-degree primary spectrum implies that $C_k$ is an $O(1)$ number at large $N$.  Consequently, the first binomial coefficient grows only as a power of $N$.  The exponential growth is controlled by the second binomial coefficient, with its largest contribution coming from $b\simeq q/k$.  Setting $q=\alpha N^2$ and using Stirling's formula gives
\begin{equation}
 \widehat d_{\alpha N^2}^{(k)}= \exp\left(B_k(\alpha)N^2+O(N)\right),
\end{equation}
where
\begin{equation}
B_k(\alpha)=\frac{1}{k} \left[(k+\alpha)\log\left(1+\frac{\alpha}{k}\right)-\alpha\log\left(\frac{\alpha}{k}\right)\right].
\end{equation}
We have therefore obtained the upper bound
\begin{equation}
 d_{\alpha N^2} < \exp\left(B_k(\alpha)N^2+O(N)\right).
\end{equation}
We can now put the two estimates together.  Multiplying the primary and secondary parts of the Hilbert series gives
\begin{equation}
 c_q^{\rm T} = d_q+\sum_{i=1}^{q}c_i^{\rm s}d_{q-i}.
\end{equation}
The coefficients $d_q$ increase with $q$.  This follows immediately from the existence of the degree-one primary invariants: multiplying a primary monomial of degree $q$ by a degree-one primary gives a primary monomial of degree $q+1$.  It follows that $d_{q-i}\leq d_q$, and hence
\begin{equation}
 c_q^{\rm T} \leq d_q\left(1+\sum_{i=1}^{q}c_i^{\rm s}\right).
\end{equation}
Therefore
\begin{equation}
 1+\sum_{i=1}^{q}c_i^{\rm s} \geq \frac{c_q^{\rm T}}{d_q}.
\end{equation}
Taking $q=\alpha N^2$ and using the lower bound on the total number of operators together with the upper bound on the primary contribution, we find
\begin{equation}
 1+\sum_{i=1}^{\alpha N^2}c_i^{\rm s} > \exp\left( \left[\alpha\log 2-B_k(\alpha)\right]N^2+O(N)\right).
\end{equation}
The left hand side counts only those secondary invariants whose degree is no larger than $\alpha N^2$.  It is therefore itself a lower bound on the total number of secondary invariants.

The explicit low-degree primary spectrum discussed in Refs.\refcite{withantal} allows us to take $k=15$.  In this case the coefficient in the exponent is
\begin{equation}
 c(\alpha) = \alpha\log 2-\frac{1}{15} \left[(15+\alpha)\log\left(1+\frac{\alpha}{15}\right)-\alpha\log\left(\frac{\alpha}{15}\right) \right].
\end{equation}
A direct examination of this function shows that
\begin{equation}
 c(\alpha)>0 \qquad\text{for}\qquad \alpha>\frac{1}{500},
\end{equation}
within the range in which the finite-$N$ cutoff does not affect the maximizing Young diagrams.  We have therefore proved that
\begin{equation}
 N_{\rm secondary} > \exp\left(c(\alpha)N^2+O(N)\right), \qquad c(\alpha)>0.
\end{equation}
Thus, the number of secondary invariants grows at least exponentially in $N^2$.  This is the careful version of the behaviour suggested by the numerical data in Table~2.

The conclusion is striking.  The number of freely acting primary generators grows only as $N^2$, but the number of elements needed to form the module basis is at least of order $e^{cN^2}$.  In the holographic interpretation, this is precisely the type of growth needed to account for a sector with an entropy proportional to $N^2$.  The primary invariants capture the perturbative degrees of freedom, while the exponentially large collection of secondary invariants has enough room to encode genuinely non-perturbative states.

\section{Physical Interpretation}

The Hironaka decomposition gives us a very clean way to organize the space of gauge invariant operators at finite $N$. Further, as we have seen in Sections \ref{onematx}, \ref{twomatx} and \ref{threeematx}, this also gives a natural decomposition of the Hilbert space of the matrix oscillator. In this decomposition, the primary invariants correspond to Fock space oscillators and hence should be identified as perturbative degrees of freedom. The secondary invariants, on the other hand, correspond to non-perturbative states and each secondary invariant supports a Fock space of excitations. 

Using the invariants as the dynamical variables leads to the collective field theory formalism~\cite{collective}. One consequence of using the invariants as dynamical degrees of freedom is that the loop expansion parameter of the resulting collective field theory is $1/N$. Consequently, it is natural to conjecture that the collective field theory constructs the dual gravitational description~\cite{constructiveholog}. In this case it is natural to identify the primary invariants with gravitons. Single trace operators in the CFT are dual to single particle states in the dual AdS spacetime. Refs.\refcite{withaniklarweh} has proved that it is always possible to choose the primary invariants to be single trace operators, which support their interpretation as perturbative degrees of freedom. On the other hand, we expect that for theories with black holes in their dual description, the black hole microstates appears as non-perturbative states, and that these states should dominate the Hilbert space at a high enough energy. There must be enough states to account for the Bekenstein-Hawking entropy
\begin{equation}
S_{BH}={A_H\over 4G_N}\sim c N^2
\end{equation}
with $c$ an order 1 number. This matches the growth in the number of secondary invariants that was established in Section \ref{CountSecondaries}.

We have used the matrix harmonic oscillator to determine the structure of the finite-$N$ singlet Hilbert space. This raises a natural question: is this description special to the free oscillator, or does it remain valid after interactions are turned on? For fixed $N$, fixed field content and the same $U(N)$ singlet constraint, turning on interactions does not change the underlying Hilbert space. It changes the Hamiltonian acting on that Hilbert space. For the matrix oscillator, the free Hamiltonian gives a particularly simple basis,
\begin{equation}
 \mathcal H_{\rm singlet} = \left\{ f\!\left(A^{1\dagger},\ldots,A^{d\dagger}\right)|0\rangle : f\ \hbox{is gauge invariant}\right\}.
\end{equation}
At finite $N$, the trace relations identify redundant invariant polynomials.
The exact singlet Hilbert space is therefore naturally identified, as a vector
space, with the invariant ring after the trace relations have been imposed.
These trace relations are matrix identities. They do not depend on the
potential or on the values of the coupling constants.

Suppose that we deform the free Hamiltonian according to
\begin{equation}
 H=H_0+gV,
\end{equation}
where $V$ is gauge invariant. Since $V$ is gauge invariant, it maps singlet states to singlet states:
\begin{equation}
 V:\mathcal H_{\rm singlet}\longrightarrow\mathcal H_{\rm singlet}.
\end{equation}
The interaction changes the energy eigenvalues and energy eigenstates, but it does not change the vector space on which the Hamiltonian acts. 

The finite-$N$ trace relations, the number and degrees of the primary invariants, the secondary invariants and the Hironaka decomposition of the invariant ring are kinematical and therefore remain unchanged when interactions are added. Consequently, one still has the exact vector-space decomposition
\begin{equation}
 \mathcal H_{\rm singlet} = \bigoplus_{\alpha} \mathbb C[p_1,\ldots,p_{N_P}]\,|\eta_\alpha\rangle,
\end{equation}
where the $p_i$ are primary invariants and the states $|\eta_\alpha\rangle$ are associated with the secondary invariants. This statement is independent of whether the Hamiltonian is free or interacting.

What does change is the relation between polynomial degree and physical energy. For the free oscillator,
\begin{equation}
 H_0\,f(A^\dagger)|0\rangle = \deg(f)\,f(A^\dagger)|0\rangle.
\end{equation}
The degree of the invariant is therefore equal to the excitation energy, and the Hilbert series is also the free oscillator partition function. After interactions are turned on, this equality is generally lost. Polynomial degree remains a useful grading of the invariant ring, but it is no longer the physical energy. Thus,
\begin{equation}
 H_{\rm Hilbert}(x) = \sum_n \dim\mathcal H_n\,x^n
\end{equation}
continues to count gauge-invariant polynomials of degree $n$, whereas the thermal partition function is
\begin{equation}
 Z(\beta) = \Tr_{\mathcal H_{\rm singlet}}\left(e^{-\beta H}\right).
\end{equation}
For an interacting Hamiltonian, these are generally different functions. The Hilbert series is determined by the kinematics, the gauge symmetry, and the finite-$N$ trace relations, and is therefore independent of the interactions. The thermal partition function depends on the interacting Hamiltonian.

The free Hamiltonian also makes the primary invariants look like independent creation operators. This simple oscillator interpretation need not be preserved by the interacting dynamics. In general, the interacting Hamiltonian can mix states with different primary occupation numbers and can also mix states built on different secondary seeds
\begin{equation}
 H\left( P_1^{n_1}\cdots P_{N_P}^{n_{N_P}}|\eta_\alpha\rangle\right) = \sum_{\beta,\{m_i\}}
 H_{\beta,\{m_i\};\alpha,\{n_i\}}\, P_1^{m_1}\cdots P_{N_P}^{m_{N_P}}|\eta_\beta\rangle.
\end{equation}
The Hironaka towers therefore remain a basis for the Hilbert space, but they need not be invariant subspaces of the interacting Hamiltonian. In particular, the Hironaka decomposition is generally neither an orthogonal decomposition nor a decomposition into interacting energy eigenspaces.

A useful analogy is provided by an ordinary particle moving in one dimension. The Hilbert space is $L^2(\mathbb R)$ whether the Hamiltonian is
\begin{equation}
 H_0=\frac{p^2}{2}+\frac{x^2}{2}\qquad {\rm or}\qquad  H=\frac{p^2}{2}+\frac{x^2}{2}+gx^4.
\end{equation}
The harmonic-oscillator states provide a basis in both cases. Turning on the interaction $gx^4$ mixes these basis states and shifts their energies, but it does not create a new Hilbert space. The matrix-model problem is analogous, except that we must first restrict to the gauge-singlet sector and impose the finite-$N$ trace relations.

The Hilbert space can genuinely change if we change the kinematical definition of the theory. Examples include changing $N$, $d$, the gauge group and the matter representations, or imposing additional conditions, such as restricting to a BPS cohomology. Changing the potential can change the spectral properties of the Hamiltonian. A confining potential may have a discrete spectrum, whereas a potential with flat directions may have a continuous spectrum. Even in this case, for fixed canonical variables and a fixed gauge constraint, the underlying singlet Hilbert space remains the same; what changes is the spectral decomposition of the Hamiltonian.

There is an additional simplification in matrix quantum mechanics. At finite $N$ and finite $d$, there are only finitely many canonical degrees of freedom. The complications associated with unitarily inequivalent Hilbert-space representations in infinite-volume quantum field theory therefore do not arise.

We conclude that the Hironaka decomposition gives an exact, interaction-independent organization of the finite-$N$ singlet Hilbert space. Interactions change the Hamiltonian, the spectrum, the energy eigenstates, and the mixing between Hironaka towers, but not the underlying vector space of gauge-invariant states. Thus, the description of the Hilbert space in terms of primary towers built on secondary seed states is completely general as an algebraic and kinematical statement. Its interpretation as a collection of independent Fock spaces is exact for the free oscillator and remains a useful organizing picture in the interacting theory, but it is not generally preserved by the interacting dynamics.

\subsection{Overcrowding}

Single traces containing fewer than $N+1$ matrices in the trace are not eliminated by the trace relations.  They must therefore be represented among the generating set of the exact finite-$N$ Hironaka decomposition.  The natural question is: are these short traces typically primary or secondary invariants? The number of primary invariants is
\begin{equation}
 h=1+(d-1)N^2,
\end{equation}
which grows as a power of $N$.  By contrast, the number of single-trace operators grows exponentially with their length.  To see this, consider a trace of $L$ matrices,
\begin{equation}
 \operatorname{Tr}\left(X^{a_1}X^{a_2}\cdots X^{a_L}\right),
 \qquad a_i\in\{1,\ldots,d\}.
\end{equation}
A crude count notices that each matrix can be any one of $d$ species, which gives a total of $d^L$ words.  This overcounts because words related by a cyclic permutation give the same trace.  Dividing by $L$ gives the useful estimate
\begin{equation}
 n_d(L)\simeq \frac{d^L}{L} =\frac{e^{L\log d}}{L}.\label{niceundercount}
\end{equation}
This is now an undercount -- because some words have a non-trivial periodicity.  For example,
\begin{equation}
 \Tr(X^1X^2X^1X^2)=\Tr(X^2X^1X^2X^1),
\end{equation}
so this word has only two distinct cyclic images rather than four.  Such exceptional words are important for obtaining the exact count, but they don't change the exponential growth.

Now consider the exact counting.  A cyclic rotation through $k$ sites fixes a word only when the letters are constant around each of the $\gcd(L,k)$ cycles produced by the rotation.  There are therefore $d^{\gcd(L,k)}$ words fixed by this rotation.  Burnside's lemma tells us to average this number over all $L$ rotations, giving
\begin{equation}
 n_d(L) =\frac{1}{L}\sum_{k=0}^{L-1}d^{\gcd(L,k)}=\frac{1}{L}\sum_{r\mid L}\varphi(r)d^{L/r},
\end{equation}
where $\varphi(r)$ is Euler's totient function.  This is the number of cyclic words, also called necklaces, of length $L$ constructed from an alphabet of $d$ letters. The first term in the divisor sum is $d^L/L$.  Every other term is exponentially smaller, so at large $L$
\begin{equation}
 n_d(L)=\frac{d^L}{L}+O_d\left(Ld^{L/2}\right).
\end{equation}
This simple estimate $d^L/L$, which we have already intuitively guessed in \eqref{niceundercount}, captures the leading growth correctly. Define
\begin{equation}
 A_d(L)=\sum_{m=1}^{L}n_d(m).
\end{equation}
This is the number of independent single-trace directions with length no larger than $L$, as long as $L\leq N$.  The restriction $L\leq N$ places us below the first trace relation, which appears at degree $N+1$.  In this stable range the different trace monomials are linearly independent. Since the sum defining $A_d(L)$ grows geometrically, it is dominated by terms close to its upper endpoint.  Using $n_d(m)\simeq d^m/m$, we obtain
\begin{align}
 A_d(L) &\simeq \sum_{m=1}^{L}\frac{d^m}{m}=\frac{d^L}{L}\left(1+\frac{1}{d}+\frac{1}{d^2}+\cdots\right)
 \left(1+O_d\left(\frac{1}{L}\right)\right) \nonumber\\
 &=\frac{d^{L+1}}{(d-1)L} \left(1+O_d\left(\frac{1}{L}\right)\right).
\end{align}
Consequently the number of loops of length at most $N$ grows exponentially,
\begin{equation}
 A_d(N)\sim \frac{d^{N+1}}{(d-1)N}.
\end{equation}
This is much faster than the $N^2$ growth of the number of primary invariants.  It follows
that, for large $N$, almost all short single-trace directions cannot be primary directions.
They must instead be carried by the secondary module.

We will now use the fact that the homogeneous system of parameters can be chosen entirely from single-trace invariants. This fact was proved in~\cite{withaniklarweh}. Fix such a choice.  Let $\mathcal T_m$ be the vector space of all single-trace invariants of degree $m$, and let $\mathcal U_m\subseteq\mathcal T_m$ be the subspace spanned by the degree-$m$ primary invariants.  Define $m_*$ to be the first degree at which not every single-trace direction has been selected as a primary,
\begin{equation}
 m_* = \min\{m\geq 1:\mathcal U_m\neq\mathcal T_m\}.
\end{equation}
Provided $m_*\leq N$, the unused part of $\mathcal T_{m_*}$ is precisely the first non-trivial secondary space. To see this, set every positive-degree primary invariant equal to zero.  In the Hironaka decomposition this operation removes all polynomial excitations of the primaries and leaves only the secondary basis.  At every degree $m<m_*$, all single-trace invariants are linear combinations of primaries.  Every multi-trace invariant is a product of such traces.  Consequently every positive-degree invariant below $m_*$ vanishes when the primaries are set to zero.

Now consider degree $m_*$.  Any genuine multi-trace operator of total degree $m_*$ is a product of traces of smaller positive degree.  Each factor is already a combination of lower-degree primaries, so every multi-trace operator again vanishes when the primaries are set to zero.  The only possible survivors are the single-trace directions in $\mathcal T_{m_*}$ that were not included in $\mathcal U_{m_*}$.  Since $m_*\leq N$, there is no trace relation that can identify one of these unused single traces with a multi-trace operator.  Thus
\begin{equation}
 \text{first secondary space}
 \simeq \frac{\mathcal T_{m_*}}{\mathcal U_{m_*}},
\end{equation}
and in particular
\begin{equation}
 \dim(\text{first secondary space})
 =\dim\mathcal T_{m_*}-\dim\mathcal U_{m_*}.
\end{equation}
This is an important point.  The first secondary does not appear because an individual trace relation has been reached.  It appears because there are already too many independent single-trace directions to fit into the finite set of primary coordinates.

\paragraph{The overcrowding scale:} There are only
\begin{equation}
 h=1+(d-1)N^2
\end{equation}
primary slots in total.  Define the overcrowding scale $L_{N,d}$ by
\begin{equation}
 L_{N,d}
 =\min\left\{L:A_d(L)>1+(d-1)N^2\right\}.
\end{equation}
By this degree there are more independent single-trace directions than there are primary slots. Every all-single-trace choice of primary invariants must omit at least one single-trace direction at or before $L_{N,d}$.  If $L_{N,d}\leq N$, the argument above shows that a non-trivial secondary invariant must appear by this degree.

We can now determine the large-$N$ behaviour of $L_{N,d}$.  Equating the asymptotic count of single traces to the number of primary directions gives
\begin{equation}
 \frac{d^{L+1}}{(d-1)L}\simeq (d-1)N^2.
\end{equation}
Equivalently,
\begin{equation}
 d^L\simeq \frac{(d-1)^2}{d}LN^2.
\end{equation}
Taking a logarithm to base $d$, we find
\begin{equation}
 L=2\log_d N+\log_d L+O_d(1).
\end{equation}
To leading order $L\simeq 2\log_d N$.  Substituting this answer back into the second logarithm gives
\begin{equation}
 L_{N,d}=2\log_d N+\log_d\log_d N+O_d(1).
\end{equation}
A secondary invariant is forced to appear at a length of order $\log N$.  Since $\log N\ll N$, this happens parametrically before the first trace relation at degree $N+1$. For fixed $d\geq2$, the stable-range condition $L_{N,d}\leq N$ is automatically satisfied at sufficiently large $N$.

Some numbers help to build intuition.  Take $d=2$ and $N=20$. There are
\begin{equation}
 h=1+20^2=401
\end{equation}
primary invariants.  Necklace counting gives $A_2(10)=261,$ and $A_2(11)=449$. Consequently, the overcrowding scale is at
\begin{equation}
 L_{20,2}=11.
\end{equation}
A secondary invariant is therefore unavoidable by degree $11$, even though the first trace relation appears only at degree $21$.  Continuing the count up to length $20$ gives $A_2(20)=111321$.
At most $401$ of these single-trace directions can be primary, so fewer than one half of one percent can belong to the primary set! Very few of the single trace invariants with a length $\le N$ are primary.

The precise degree at which the first secondary invariant appears depends on the choice of primary coordinates.  A
poor choice can omit a low-degree single-trace direction and make a secondary appear earlier. The overcrowding scale is therefore an upper bound on how long the appearance of a secondary can be postponed.  For the case that we intentionally choose each primary invariant to assume the lowest possible degree, the overcrowding scale gives us a fair idea of where the first secondary appears. What is independent of the choice of primaries is the capacity statement: no all-single-trace homogeneous system of parameters can accommodate every single-trace
direction beyond $L_{N,d}$.

It is also important not to conclude that every primary invariant must have length no larger than $N$.  Consider, for example, the two-matrix model with $N=3$.  Its graded partition function can be written as
\begin{align}
 Z(x,y) &=\frac{1+x^3y^3}{(1-x)(1-y)(1-x^2)(1-xy)(1-y^2)(1-x^3)}\nonumber\\
 &\hspace{1cm}\times \frac{1} {(1-x^2y)(1-xy^2)(1-y^3)(1-x^2y^2)}.
\end{align}
The factor $1-x^2y^2$ in the denominator corresponds to a primary invariant of total length four, which may be chosen to be
\begin{equation}
 \operatorname{Tr}(X^2Y^2).
\end{equation}
Thus it is not unusual for a small number of primary invariants to have length larger than $N$.  Overcrowding
does not impose a sharp upper bound on the length of an individual primary.  It constrains the total number of independent single-trace directions that can simultaneously be used as primary coordinates.

There are a number of natural scales that are associated with finite $N$ physics: the first trace relation~\cite{procesi} appears at degree $L_{\rm trace}=N+1$ and it indicates that certain operators are no longer linearly independent~\cite{SEP}. Another physically motivated scale is the scale $L_{\rm fact}=O(N\log N)$ where large-$N$ factorization of gauge invariant correlators starts to fail~\cite{failfact}. The overcrowding scale $L_{N,d}=O(\log N)$ is parameterically below both of these scales. It is the scale at which it is no longer possible to promote every independent single trace coordinate to a freely acting coordinate of the finite-$N$ invariant ring. Overcrowding is an algebraic precursor of finite-$N$ physics: the finite capacity of the perturbative coordinate space is visible long before individual trace relations become active.

There is a physical consequence of overcrowding that makes contact with the physics of fast scrambling~\cite{fastscramblers}. The logarithmic form of the overcrowding scale is the same as the fast-scrambling time. This is no accident: the two counting arguments are naturally related. There are, however, important differences too
\begin{itemize} 
\item Overcrowding is a kinematical statement: it follows from the invariant ring and does not depend on the
Hamiltonian.  
\item Scrambling is a dynamical statement, made for a specific Hamiltonian.
\end{itemize}
Consider a chaotic matrix model whose Hamiltonian is constructed from traces of words of bounded degree $r$.  A commutator with the Hamiltonian can increase the word length of an operator by only a bounded amount.  Schematically,
\begin{equation}
 \deg[H,\mathcal O]\leq \deg\mathcal O+r-2.
\end{equation}
After the action of $n$ nested commutators,
\begin{equation}
 \deg\left(\operatorname{ad}_H^n\mathcal O\right) \lesssim L_0+n(r-2),
\end{equation}
where $L_0$ is the initial length.  The Heisenberg operator
\begin{equation}
 \mathcal O(t) =\sum_{n=0}^{\infty}\frac{(it)^n}{n!}{\rm ad}_H^n\mathcal O
\end{equation}
therefore has an operator front whose maximal possible length grows at most linearly with time. There is numerical evidence that supports a characteristic word length growing ballistically~\cite{ballistic}
\begin{equation}
 L(t)\simeq v_{\rm op}t.
\end{equation}
The number of distinct single-trace directions available behind this front grows exponentially,
\begin{equation}
 K(t)\sim d^{L(t)}\sim e^{\lambda_{\rm op}t}, \qquad \lambda_{\rm op}=v_{\rm op}\log d.
\end{equation}
The primary invariants provide the independent perturbative directions, whose number is
\begin{equation}
 N_{\rm pert}=1+(d-1)N^2\sim N^2.
\end{equation}
Suppose that a simple single-trace operator creates a perturbation which initially occupies only one or a few of these directions.  Under chaotic evolution the perturbation spreads through an increasing number of primary directions.  Scrambling occurs when recovering the information requires access to an order-one fraction of the complete perturbative sector. Thus the scrambling time $t_*$ is determined by
\begin{equation}
 e^{\lambda_{\rm op}t_*}\sim N_{\rm pert}\sim N^2,
\end{equation}
and hence
\begin{equation}
 t_*\sim \frac{2}{\lambda_{\rm op}}\log N+O(1).
\end{equation}
This is the usual fast-scrambling scale. The same answer follows directly from overcrowding.  The operator has enough available single-trace components to explore the full perturbative sector when
\begin{equation}
 A_d(L(t_*))\sim N_{\rm pert}.
\end{equation}
By definition this occurs when $L(t_*)\simeq L_{N,d}$.  Using ballistic growth gives
\begin{align}
 t_* &\simeq \frac{L_{N,d}}{v_{\rm op}}=\frac{1}{v_{\rm op}} \left(2\log_dN+\log_d\log_dN+O_d(1)\right)\nonumber\\
 &=\frac{2\log N+\log\log N+O(1)}{\lambda_{\rm op}}.
\end{align}
The derivation of the overcrowding scale and the derivation of the scrambling time are therefore almost identical
\begin{align}
 \text{overcrowding:}\qquad &A_d(L_{N,d})\sim N_{\rm pert},\qquad A_d(L_{N,d})\sim \frac{d^{L_{N,d}}}{L_{N,d}}\sim e^{L_{N,d}\log d} \nonumber\\
 \text{scrambling:}\qquad &K(t_*)\sim N_{\rm pert},\qquad \qquad K(t_*)=e^{v_{\rm op}t_*} \nonumber\\
 \text{ballistic growth:}\qquad &L(t_*)=v_{\rm op}t_*\sim L_{N,d}.
\end{align}
Under these assumptions, overcrowding supplies a microscopic algebraic picture of the onset of scrambling.  It is the point at which a growing operator has enough distinct single-trace components to explore an order-one fraction of all perturbative directions.

There are two qualifications.  Invariant theory alone does not prove that a given Hamiltonian scrambles.  A free or integrable Hamiltonian may fail to spread an operator democratically through the available directions.  Second, the primary invariants are algebraically independent coordinates, but they are not automatically orthogonal subsystems or qubits.  A precise dynamical statement requires an inner product, supplied naturally by matrix-model correlation functions, together with control over the relevant energy sector. The robust algebraic conclusion is that the perturbative sector has only $O(N^2)$ independent directions and that the exponentially proliferating single traces fill this capacity at length $O(\log N)$.  When chaotic dynamics explores these directions ballistically and sufficiently democratically, this algebraic overcrowding becomes the shadow of fast scrambling.

\subsection{Secondary invariants as discrete redundancy}\label{SecAsDisc}

In this section we will build more insight into the interpretation of the primary and secondary invariants. The primary invariants are continuous coordinates on the space of gauge invariants, while the secondary invariants provide a discrete amount of additional information. We will illustrate the argument in the context of the $d=4$ matrix model with $N=2$.

Our Hermitian $2\times2$ matrices can always be expanded as ($a=1,2,3,4$, $\vec{x}_a\in{\mathbb R}^3$)
\begin{equation}
X^a={t_a\over 2}{\bf 1}+\vec x_a\cdot\vec\sigma,\qquad{\rm where}\qquad t_a=\tr X_a.\label{newvariables}
\end{equation}
The original gauge invariance, which acted as simultaneous $U(2)$ conjugation $X^a\to UX^aU^\dagger$ leaves the $t_a$ invariant and acts on the $\vec{x}_a$ as a common rotation:
\begin{equation}
\vec x_a\longmapsto R(U)\vec x_a,\qquad R(U)\in SO(3).
\end{equation}
Thus, the problem of constructing all gauge invariants from the four matrices reduces to the problem of computing the invariants under simultaneous rotations of the vectors $\vec{x}_a$. These invariants, together with the $t^a$ furnish the complete set of gauge invariants. The most obvious set of invariants constructed out of the vectors is set of ten independent dot products of the vectors $\vec{x}_a$
\begin{equation}
Q_{ab}=\vec x_a\cdot\vec x_b,\qquad 1\le a\le b\le 4.\label{scalarprods}
\end{equation}
The Gram matrix $G$ is the matrix obtained by assembling these dot products into a matrix. Since four vectors in ${\mathbb R}^3$ are linearly dependent, the determinant of the Gram matrix $G$ vanishes 
\begin{equation}
G=(Q_{ab}),\qquad\qquad\det G=0.
\end{equation}
There is another class of invariants, given by the scalar triple products
\begin{equation}
T_{abc}=\vec x_a\cdot(\vec x_b\times \vec x_c).
\end{equation}

To construct the Hironaka decomposition of the ring, we want to construct a complete set of primary and secondary invariants. It is helpful to note that the Hilbert series for $N=2$, $d=4$ is
\begin{equation}
H_{2,4}(q)={1+q^2+4q^3+q^4+q^6\over (1-q)^4(1-q^2)^9}.\label{LstHlbSr}
\end{equation}
Thus, we can immediately read off:
\begin{equation}
\begin{array}{ccl}
(1-q)^{-4} &\Rightarrow& 4\text{ degree-one primaries},\\[1mm]
(1-q^2)^{-9} &\Rightarrow& 9\text{ degree-two primaries},\\[1mm]
1+q^2+4q^3+q^4+q^6 &\Rightarrow& 8\text{ secondaries.}
\end{array}\nonumber
\end{equation}
The choice of the 4 degree 1 primaries is obvious
\begin{equation}
 p_1=t_1, \quad p_2=t_2, \quad p_3=t_3, \quad p_4=t_4,
\end{equation}
The 9 degree 2 primaries must be selected from the 10 independent scalar products \eqref{scalarprods}. A suitable choice is given by 
\begin{equation}
 p_5=Q_{11},\quad p_6=Q_{22},\quad p_7=Q_{33},\quad p_8=Q_{44},\quad  p_9=Q_{13},
\end{equation}
\begin{equation}
p_{10}=Q_{14},\quad p_{11}=Q_{23},\quad p_{12}=Q_{24},\quad p_{13}=Q_{12}-Q_{34}.
\end{equation}
The remaining quadratic direction is
\begin{equation}
\Sigma={1\over2}(Q_{12}+Q_{34}).
\end{equation}
In terms of primaries and $\Sigma$, we have $Q_{12}=\Sigma+{p_{13}\over2}$ and $Q_{34}=\Sigma-{p_{13}\over2}$. The Gram matrix is
\begin{equation}
G(p,\Sigma)=
\begin{pmatrix}
 p_5 & \Sigma+\frac12p_{13} & p_9 & p_{10}\\
 \Sigma+\frac12p_{13} & p_6 & p_{11} & p_{12}\\
 p_9 & p_{11} & p_7 & \Sigma-\frac12p_{13}\\
 p_{10} & p_{12} & \Sigma-\frac12p_{13} & p_8
\end{pmatrix}.
\end{equation}
Thus the fact that the determinant of the Gram matrix vanishes gives the following quartic equation for $\Sigma$
\begin{equation}
\det G(p,\Sigma)=\Sigma^4+f_1(p_i)\Sigma^3+f_2(p_i)\Sigma^2+f_3(p_i)\Sigma+f_4(p_i)=0,\label{QrtcEqn}
\end{equation}
where the $f_i(p)$ are polynomials in the primary invariants that are simply worked out by computing $\det G(p,\Sigma)$ explicitly. Notice that even for fixed values of the primary invariants $p_i$, there are generically four complex roots for $\Sigma$
\begin{equation}
\Sigma\in\{\Sigma_1(p),\Sigma_2(p),\Sigma_3(p),\Sigma_4(p)\}.
\end{equation}
This is not the only ambiguity that remains after fixing all the primary invariants. Notice that if we perform a sign inversion on all the $\vec{x}_a$ i.e. we send $\vec{x}_a\to-\vec{x}_a$, the inner products $Q_{ab}$ are left invariant, but the scalar triple products all change sign. This implies that, even after we know the primaries we still do not know the signs of the scalar triple product. Explicit computation shows that the ambiguity is a single overall sign
\begin{equation}
\{\epsilon T_{123},\,\,\epsilon T_{124},\,\,\epsilon T_{134},\,\,\epsilon T_{234}\},\,\,\qquad \epsilon=\pm1.
\end{equation}
A complete secondary basis is given by $\eta_0=1$ and
\begin{equation}
\eta_1=\Sigma,\quad\eta_2=\Sigma^2,\quad\eta_3=\Sigma^3,\quad\eta_4=T_{123},\quad \eta_5=T_{124},\quad\eta_6=T_{134},\quad\eta_7=T_{234}.
\end{equation}
Notice that the degrees of these secondary invariants are $0,2,4,6,3,3,3,3$ which matches the prediction of the numerator of the Hilbert series \eqref{LstHlbSr}. An important property that any list of secondary invariants must reproduce, is the product rule given in \eqref{SecProdRule}. The product rule has a few ingredients: First, the quartic \eqref{QrtcEqn} reduces high powers of $\eta_1=\Sigma$, so that, for example
\begin{equation}
(\eta_2)^2=-f_1(p)\eta_3-f_2(p)\eta_2-f_3(p)\eta_1-f_4(p)\eta_0.\label{eta2eta2}
\end{equation}
All products of $\eta_1,\eta_2$ and $\eta_3$ can be reduced using the definition of the $\eta$'s, using \eqref{QrtcEqn} and simple algebra. Products of cubic secondaries are reduced using Gram matrix minors
\begin{equation}
T_{ijk}T_{\ell mn}=\det\big(Q_{ab}\big)_{a\in\{i,j,k\},\;b\in\{\ell,m,n\}}.\nonumber
\end{equation}
Finally, to reduce products of cubic invariants and powers of $\Sigma$ the key observation is that because the four vectors $\vec x_1,\ldots,\vec x_4$ lie in three dimensions, they satisfy the linear-dependence identity
\begin{equation}
 T_{234}\vec x_1 -T_{134}\vec x_2 +T_{124}\vec x_3 -T_{123}\vec x_4 =0. \label{eq:vector-identity}
\end{equation}
Taking scalar products of this identity with $\vec{x}_a$ supplies the final identities we need to prove that the secondary invariants are quadratically reducible. An example of a product rule is
\begin{equation}
\eta_1\eta_4=\Sigma T_{123}={p_{13}\over2}\eta_4+p_7\eta_5-p_{11}\eta_6+p_9\eta_7.\nonumber
\end{equation}

Choose values for all of the primary invariants $p_i$, $i=1,\cdots,13$. This does not specify the complete set of gauge invariant operators. Indeed, $\Sigma$ must be chosen from one of four roots of a quartic equation
\begin{equation}
\Sigma=\Sigma_r(p),\qquad r=1,2,3,4.\nonumber
\end{equation}
For each choice of root, we must still choose one common overall sign for the cubic invariants
\begin{equation}
T_I\longmapsto -T_I,\qquad I\in\{123,124,134,234\}.\nonumber
\end{equation}
Therefore the complexified quotient has generic fiber
\begin{equation}
\{(p_k,\Sigma_r(p_k),\epsilon):\ r=1,2,3,4,\quad \epsilon=\pm1\}\nonumber
\end{equation}
and hence is an eight-sheeted algebraic cover of primary space.

Not all four choices of the root for $\Sigma$ are always possible. Since our matrices are Hermitian, we know that the Gram matrix is positive semi-definite $G\succeq0$. Not every root of \eqref{QrtcEqn} gives a positive semi-definite Gram matrix $G$ and so not every root is a valid choice. To explore this explicitly, choose the point in the primary base given by
\begin{eqnarray}
&&Q_{11}=1,\quad Q_{22}=2,\quad Q_{33}=1,\quad Q_{44}=2,\quad Q_{24}=1,\nonumber\\
&&\qquad Q_{13}=Q_{14}=Q_{23}=0,\qquad Q_{12}-Q_{34}=0.\label{ExPPrimChoice}
\end{eqnarray}
At this point $Q_{12}=Q_{34}=\Sigma$ and
\begin{equation}
G(\Sigma)=
\begin{pmatrix}
1&\Sigma&0&0\\   \Sigma&2&0&1\\
0&0&1&\Sigma\\    0&1&\Sigma&2
\end{pmatrix},
\qquad\Rightarrow\qquad  \det G(\Sigma)=(\Sigma^2-1)(\Sigma^2-3)=0.
\end{equation}
The four roots for $\Sigma$ are given by
\begin{equation}
\Sigma=\pm1,\qquad \Sigma=\pm\sqrt3.
\end{equation}
If a matrix is positive semi-definite, then all of its principal minors are also positive semi-definite. The roots $\pm\sqrt3$ are not physical because the principal minor obtained by restricting to the first and second rows and columns is negative
\begin{equation}
\det\begin{pmatrix}Q_{11}&Q_{12}\\Q_{12}&Q_{22}\end{pmatrix}=2-\Sigma^2=-1<0.
\end{equation}
Thus, at this point in the primary base, the physical fiber is
\begin{equation}
(s,\epsilon)\in\{+1,-1\}\times\{+1,-1\},\qquad \Sigma=s.\label{ActFiber}
\end{equation}
The picture that emerges is that our space of gauge invariant operators is described by an algebraic curve over the base provided by the primary invariants. There are a total of 8 sheets in the algebraic cover (4 roots$\times$2 signs) which matches the fact that we have 8 secondary invariants. In this way of thinking about things, the secondary invariants parametrise 8 discrete data points which are left over after the continuous primary invariants have been specified. 

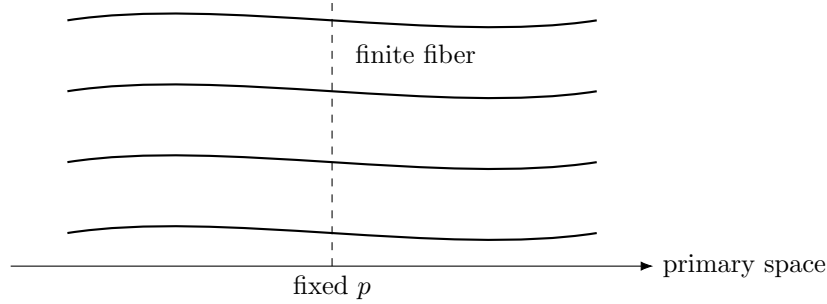
\begin{center}
\begin{tikzpicture}[>=Latex,scale=1.25]
  \draw[->] (-3.4,-1.5)--(3.4,-1.5) node[right] {primary space};
  \foreach \y/\c in {1.1/black,0.35/black,-0.4/black,-1.15/black}{
    \draw[thick,\c] (-2.8,\y) .. controls (-1.2,\y+0.25) and (1.2,\y-0.25) .. (2.8,\y);
  }
  \draw[dashed] (0,-1.5)--(0,1.35);
  \node[right] at (0.15,0.75) {finite fiber};
  \node[below] at (0,-1.5) {fixed $p$};
\end{tikzpicture}
\end{center}
\begin{quote}
The space of gauge invariant operators is an eight-sheeted algebraic cover of primary space. The horizontal line stands for the 13-dimensional primary space. The different curved surfaces above the primary base are distinguished by the values of the secondary invariants.
\end{quote}

\subsection{Path Integral quantization}

Our physical interpretation of the primary and secondary invariants has been motivated by the decomposition of the Hilbert space of the matrix oscillator achieved by the Hironaka decomposition. An alternative to quantizing a theory using operators acting on a Hilbert space is to use the path integral. Is the physical interpretation that we have suggested also consistent with a path integral quantization of the theory?

In a path integral quantization, we evaluate the functional integral given by
\begin{equation}
Z=\int [d\phi] e^{\frac{i}{\hbar}S}\label{pathintegral}
\end{equation}
Correlation functions of fields are given by
\begin{equation}
\langle\phi(x_1)\cdots\phi(x_n)\rangle=Z^{-1}\int [d\phi] e^{\frac{i}{\hbar}S}\phi(x_1)\cdots\phi(x_n)
\end{equation}
With the exception of free theories, which reduce to Gaussian integrals, the path integral \eqref{pathintegral} is typically too difficult to evaluate and we must consider some approximation. In the semiclassical limit where $\hbar$ is small, the integrand of \eqref{pathintegral} oscillates wildly and the contribution from most field configurations cancel each other out. An important exception is when we are close to a classical saddle. At the classical saddle the classical equation of motion is satisfied and we have
\begin{equation}
{\delta S\over\delta\phi}=0.\label{statephase}
\end{equation}
This equation \eqref{statephase} is nothing but the statement that the phase of the path integral is stationary. Consequently, near these classical saddles the phase becomes stationary and we obtain a large contribution to the path integral. We can evaluate the path integral in the semiclassical limit by summing over each of these saddles and integrating over the small fluctuations about each saddle. The result is~\cite{colemansemiclassical}
\begin{equation}
Z=\sum_{k\in{\rm saddles}}e^{\frac{i}{\hbar}S[\phi^{(k)}]}\int [d\eta] e^{\frac{i}{\hbar}S_{\rm fluct}[\eta]}\label{semiclassical}
\end{equation}
where $\eta$ denotes the small fluctuation about the classical saddle. The path integral over $\eta$ can be performed as usual, by deriving a set of Feynman rules and performing a loop expansion.

\noindent
The approximation \eqref{semiclassical} has an interesting interpretation. First note that this is non-perturbative in $\hbar$ because the factors $e^{\frac{i}{\hbar}S[\phi^{(k)}]}$ do not admit a power series expansion in $\hbar$. Consequently, the formula \eqref{semiclassical} is performing a sum over non-perturbative states. In a theory with soliton solutions, for example, each saddle may be a configuration with a definite number of solitons. Each such saddle can be excited by an arbitrary number of perturbative excitations. The contribution of these perturbative excitations is captured in the path integral language by performing an integral over the fluctuation $\eta$. It is clear that in the path integral formalism there is a sharp distinction between non-perturbative states and perturbative degrees of freedom: we integrate over perturbative degrees of freedom and sum over non-perturbative states.

The simplest analogue of the path integral arises if we consider field theory in $0+0$ dimensions. In $d+1$ dimensions we are used to writing the field as $\phi(t,\vec{x})$. This field will associate a (real or complex) number to each point $(t,\vec{x})$ in the $d+1$ dimensional spacetime. Since there is only a single point in $0+0$ dimensions, the field is now simply a (real or complex) number and the path integral reduces to an ordinary integral. We can play all of the field theory tricks we like with this ordinary integral, including defining Feynman rules and carrying out the loop expansion. The path integral for matrix theories in $0+0$ dimensions reduces to an ordinary matrix integral and for this problem we can see the consequences of the Hironaka decomposition explicitly. We will now study matrix integrals and show they can be written as follows~\cite{withjoao}
\begin{equation}
Z=\int \prod_{a=1}^d[dX^a]e^{-S}=\sum_{k\in{\rm secondary\,\, sectors}}\int \prod_{a=1}^h dp_a e^{-S^{(k)}}\label{PIforHironaka}
\end{equation} 
The integral is over all primary invariants, so that the primary invariants do indeed appear as expected of perturbative degrees of freedom. The sum runs over a number of sectors, with the number of sectors equal to the number of secondary invariants, supporting the idea that the secondary invariants are associated with non-perturbative states of the theory. The values of the secondary invariants identify the different sectors, but it is incorrect to assume that each sector is literally associated to a particular secondary invariant. 

Why should a secondary invariant be thought of as non-perturbative? The answer comes from the geometry of
the invariant variables themselves. The primary invariants provide continuous coordinates, while the secondary invariants distinguish a finite number of different algebraic branches lying above the same values of the primaries. A perturbative expansion is local and therefore probes only one such branch at a time. The secondary invariants remember the additional global information which is invisible to a purely local perturbative expansion. A particularly instructive example is provided by the integral over four $2\times 2$ Hermitian matrices. In this example the geometry is rich enough to display all of the key ideas, while still being simple enough that everything can be described explicitly. Towards this end, consider the matrix integral
\begin{equation}
 Z=\int\prod_{a=1}^{4}[dX^a]\,e^{-S(X^1,X^2,X^3,X^4)} .
\end{equation}
This is an ordinary integral over $4N^2=16$ variables. We assume that the action is invariant under simultaneous $U(2)$ conjugation. Our goal is not to evaluate this integral for a particular choice of $S$. Instead, we want to change variables from the matrix entries to the invariant variables and see what becomes of the path integral.

The first step is immediate from the decomposition \eqref{newvariables}.  The change from the four matrix entries of $X^a$ to $t_a$ and $\vec{x}_a$ is linear and has unit Jacobian, so that
\begin{equation}
 \prod_{a=1}^{4}[dX^a]= \prod_{a=1}^{4}dt_a\, \prod_{a=1}^{4}d^3x_a .
\end{equation}
The $t_a$ are already the first four primary invariants.  The non-trivial step is to rewrite the integral over the four vectors $\vec{x}_a$ in terms of the scalar products $Q_{ab}$. We do this on a generic patch on which $\vec{x}_1,\vec{x}_2,\vec{x}_3$ are linearly independent.  Choose an orthonormal frame $(\hat e_1,\hat e_2,\hat e_3)$ and write
\begin{equation}
 \vec{x}_1=r_1\hat e_1,\qquad \vec{x}_2=a\hat e_1+b\hat e_2,\qquad
 \vec{x}_3=c\hat e_1+d\hat e_2+\epsilon e\hat e_3, \qquad \epsilon=\pm1, \label{eq:adapted-four-vectors}
\end{equation}
with $r_1,b,e>0$.  The sign $\epsilon$ is the orientation sign of the cubic invariants.  The common orientation of the frame is pure gauge and is parametrized by an element of $SO(3)$. Introduce the $3\times3$ Gram matrix of the first three vectors,
\begin{equation}
 A= \begin{pmatrix} Q_{11}&Q_{12}&Q_{13}\\ Q_{12}&Q_{22}&Q_{23}\\ Q_{13}&Q_{23}&Q_{33}\end{pmatrix}.
\end{equation}
Using (\ref{eq:adapted-four-vectors}) we easily find
\begin{equation}
 \begin{split}
 Q_{11}&=r_1^2,\qquad Q_{12}=r_1a,\qquad Q_{13}=r_1c,\\
 Q_{22}&=a^2+b^2,\qquad Q_{23}=ac+bd,\qquad Q_{33}=c^2+d^2+e^2 .
 \end{split}
\end{equation}
A direct computation gives
\begin{equation}
 \left| {\partial(Q_{11},Q_{12},Q_{13},Q_{22},Q_{23},Q_{33})  \over \partial(r_1,a,b,c,d,e)}\right|=8r_1^3b^2e .
\end{equation}
On the other hand, $\det A=r_1^2b^2e^2$. Clearly, for a fixed value of the orientation sign,
\begin{equation}
 d^3x_1\,d^3x_2\,d^3x_3 = {1\over8\sqrt{\det A}}\, d^6A\,d\mu_{SO(3)} ,
\end{equation}
where $d^6A$ denotes the measure over the six independent entries of $A$. We now include the fourth vector.  Since the first three vectors form a basis on our patch, we can write
\begin{equation}
 \vec{x}_4 = c_1\vec{x}_1+c_2\vec{x}_2+c_3\vec{x}_3 .
\end{equation}
Define
\begin{equation}
\vec{\nu}=\begin{pmatrix} Q_{14}\\ Q_{24}\\ Q_{34} \end{pmatrix}, \qquad
\vec{c}= \begin{pmatrix} c_1\\c_2\\c_3 \end{pmatrix}.
\end{equation}
Taking scalar products with the first three vectors gives $\vec{\nu}=A\vec{c}$. Further,
\begin{equation}
 Q_{44}=\vec{c}^{\,T}A\vec{c}=\vec{\nu}^T A^{-1}\vec{\nu}.
\end{equation}
The volume of the parallelepiped spanned by $\vec{x}_1,\vec{x}_2,\vec{x}_3$ is $\sqrt{\det A}$, and hence
\begin{equation}
 d^3x_4 = \sqrt{\det A}\,d^3 c = {d^3\nu\over\sqrt{\det A}} .
\end{equation}
Multiplying the two measures we obtain
\begin{equation}
 \prod_{a=1}^{4}d^3x_a = {d^6A\,d^3\nu\,d\mu_{SO(3)}\over8\det A}
\end{equation}
for each choice of $\epsilon$.  The full Gram matrix has the block form
\begin{equation}
 G= \begin{pmatrix} A&\nu\\ \nu^T&Q_{44} \end{pmatrix}\qquad\Rightarrow\qquad
 \det G = (\det A)\left(Q_{44}-\nu^TA^{-1}\nu\right).
\end{equation}
Therefore
\begin{equation}
 \delta(\det G) = {1\over\det A}\, \delta\left(Q_{44}-\nu^TA^{-1}\nu\right),
\end{equation}
and the vector measure becomes
\begin{equation}
 \prod_{a=1}^{4}d^3x_a= {1\over8} \left(\prod_{1\leq a\leq b\leq4}dQ_{ab}\right) d\mu_{SO(3)}\, \Theta(G\succeq0)\,
 \delta(\det G). \label{eq:four-vector-measure}
\end{equation}
The factor $\Theta(G\succeq0)$ is needed because a real symmetric matrix is a Gram matrix of real vectors only when it is positive semi-definite. The integral over the common $SO(3)$ orientation is now trivial.  With
\begin{equation}
 \int d\mu_{SO(3)}=8\pi^2,
\end{equation}
equation (\ref{eq:four-vector-measure}) gives
\begin{equation}
 Z = \pi^2\sum_{\epsilon=\pm1}\int\left(\prod_{a=1}^{4}dt_a\right) \left(\prod_{1\leq a\leq b\leq4}dQ_{ab}\right)
 \Theta(G\succeq0)\, \delta(\det G)\, e^{-S_\epsilon(t,Q)} . \label{eq:path-integral-Q}
\end{equation}
The subscript $\epsilon$ reminds us that if the action depends on the cubic invariants, their common sign must also be specified.  If the action depends only on the $Q_{ab}$, the two orientations give identical contributions.

We now use exactly the primary variables chosen in Section \ref{SecAsDisc}. The only non-trivial part of the change of variables is $p_{13}=Q_{12}-Q_{34}$ and $\Sigma={1\over2}(Q_{12}+Q_{34})$, for which
\begin{equation}
 \left| {\partial(p_{13},\Sigma)\over\partial(Q_{12},Q_{34})}\right|=1.
\end{equation}
All the remaining quadratic variables are already primary invariants. Consequently,
\begin{equation}
 \left(\prod_{a=1}^{4}dt_a\right) \left(\prod_{1\leq a\leq b\leq4}dQ_{ab}\right)= d^{13}p\,d\Sigma .
\end{equation}
Using the notation of \eqref{QrtcEqn}, define
\begin{equation}
 \Phi(p,\Sigma)\equiv\det G(p,\Sigma)= \Sigma^4+f_1(p)\Sigma^3+f_2(p)\Sigma^2+f_3(p)\Sigma+f_4(p).
\end{equation}
Equation (\ref{eq:path-integral-Q}) therefore becomes
\begin{equation}
 Z = \pi^2\sum_{\epsilon=\pm1} \int d^{13}p\,d\Sigma\, \Theta(G(p,\Sigma)\succeq0)\,
 \delta\bigl(\Phi(p,\Sigma)\bigr)\, e^{-S_\epsilon(p,\Sigma)} . \label{eq:path-integral-primary-sigma}
\end{equation}
This formula makes the role of the quartic equation transparent. On the generic locus the four roots are distinct, and we can use
\begin{equation}
 \delta\bigl(\Phi(p,\Sigma)\bigr)=\sum_{r=1}^{4}{\delta\bigl(\Sigma-\Sigma_r(p)\bigr)\over
  \left|\partial_\Sigma\Phi(p,\Sigma_r(p))\right|}. \label{eq:quartic-delta}
\end{equation}
Using the delta function to perform the $\Sigma$ integral, we find
\begin{equation}
 Z=\sum_{r=1}^{4}\sum_{\epsilon=\pm1} \,\,\pi^2\int d^{13}p {\Theta_r(p)\over\left|\partial_\Sigma\Phi(p,\Sigma_r(p))\right|}e^{-S_{r,\epsilon}(p)},\label{eq:branchwise-four-matrix-integral}
\end{equation}
where, as usual
\begin{equation}
 \Theta_r(p)= \begin{cases}1,&\Sigma_r(p)\in{\mathbb R} \ \hbox{and}\ G(p,\Sigma_r(p))\succeq0,\\
 0,&\hbox{otherwise}. \end{cases}
\end{equation}
The notation $S_{r,\epsilon}(p)$ means that the action is evaluated after setting $\Sigma=\Sigma_r(p)$ and choosing the common orientation sign $\epsilon$ for the four cubic invariants.

Equation (\ref{eq:branchwise-four-matrix-integral}) is the precise version of the schematic formula \eqref{PIforHironaka}.  The thirteen primary invariants are genuine continuous integration variables.  By contrast, $r$ and $\epsilon$ are discrete labels and are summed over.  Thus the distinction suggested by the Hironaka decomposition appears directly in the integral: primaries are integrated over and secondary data are summed over. Notice that no semiclassical approximation has been made in deriving this result.  The sum over sectors is already present in the exact change of variables to gauge invariant coordinates.

There is a useful way to understand why the secondary invariants appear as sector data rather than as extra integration variables.  Introduce $T=(T_{123},T_{124},T_{134},T_{234})$. The product identity discussed above in \eqref{eta2eta2} implies that the cubic invariants obey polynomial constraints.  A convenient set of constraints is
\begin{equation}
 \begin{split}
 F_1&= T_{123}^2 - \det(Q_{ab})_{a,b\in\{1,2,3\}},\qquad\,\,\,\,\,\,\,\,
 F_2= T_{123}T_{124} - \det(Q_{ab})_{\substack{a\in\{1,2,3\}\\b\in\{1,2,4\}}},\\
 F_3&= T_{123}T_{134} - \det(Q_{ab})_{\substack{a\in\{1,2,3\}\\b\in\{1,3,4\}}},\qquad
 F_4= T_{123}T_{234} - \det(Q_{ab})_{\substack{a\in\{1,2,3\}\\b\in\{2,3,4\}}}.
 \end{split} \label{eq:cubic-constraints}
\end{equation}
On the generic locus, fixing $p_i$ and one of the roots $\Sigma_r(p)$ leaves exactly two solutions of $F_A=0$, $A=1,2,3,4,$ and these are precisely the two choices $\epsilon=\pm1$. This observation allows us to combine the discrete sum into a single integral. Define
\begin{equation}
 {\cal J}_T = \left| \det{\partial(F_1,F_2,F_3,F_4) \over \partial(T_{123},T_{124},T_{134},T_{234})} \right|.
\end{equation}
For the choice (\ref{eq:cubic-constraints}) we find ${\cal J}_T=2|T_{123}|^4$. The elementary multidimensional delta-function identity then gives
\begin{equation}
 {\cal J}_T\prod_{A=1}^{4}\delta(F_A) = \sum_{\epsilon=\pm1} \delta^{(4)}\bigl(T-T^{(\epsilon)}(p,\Sigma)\bigr).
\end{equation}
Consequently, the exact integral can equally well be written as
\begin{equation}
 Z=\pi^2\int d^{13}p\,d\Sigma\,d^4T\, \Theta(G(p,\Sigma)\succeq0)\, \delta\bigl(\Phi(p,\Sigma)\bigr)\,
 {\cal J}_T \prod_{A=1}^{4}\delta(F_A)\,\, e^{-S(p,\Sigma,T)} . \label{eq:single-invariant-integral}
\end{equation}
Integrating first over $T$ reproduces the sum over the two orientation signs. Integrating next over $\Sigma$ reproduces the sum over the four roots in (\ref{eq:branchwise-four-matrix-integral}).  Thus (\ref{eq:single-invariant-integral}) and (\ref{eq:branchwise-four-matrix-integral}) are simply two different ways of writing the same matrix integral.

There is now a striking connection with the saddle-point discussion at the beginning of this section.  Write the delta functions in Fourier form by introducing Lagrange multipliers $\lambda$ and $\mu_A$.  The part of the effective action involving the auxiliary variables is
\begin{equation}
 S_{\rm eff} = S(p,\Sigma,T)-\log{\cal J}_T -i\lambda\,\Phi(p,\Sigma) -i\sum_{A=1}^{4}\mu_A F_A(p,\Sigma,T).
 \label{eq:invariant-effective-action}
\end{equation}
The equations obtained by varying the Lagrange multipliers are
\begin{equation}
 \Phi(p,\Sigma)=0, \qquad F_A(p,\Sigma,T)=0, \qquad A=1,2,3,4. \label{eq:branch-saddle-equations}
\end{equation}
But these are exactly the equations that define the algebraic branches found in Section \ref{SecAsDisc}. To see that these really are stationary points of the enlarged integral, consider a generic point for which
\begin{equation}
 \partial_\Sigma\Phi\neq0, \qquad \det{\partial F_A\over\partial T_I}\neq0.
\end{equation}
The four equations obtained by varying the $T_I$ determine the four multipliers $\mu_A$.  Once these have been fixed, the equation obtained by varying $\Sigma$ determines $\lambda$, since $\partial_\Sigma\Phi\neq0$.  Thus every generic solution
\begin{equation}
 \bigl(\Sigma_r(p),T^{(r,\epsilon)}(p)\bigr), \qquad r=1,2,3,4,\qquad \epsilon=\pm1,
\end{equation}
lifts to a stationary point of one and the same invariant-space action. This statement needs one important qualification. The auxiliary variational problem naturally sees the full complexified algebraic cover.  The original integral, however, is over Hermitian matrices and therefore retains only those solutions for which
\begin{equation}
 G(p,\Sigma_r(p))\succeq0.
\end{equation}
As we explicitly saw in \eqref{ExPPrimChoice}--\eqref{ActFiber}, some algebraic roots can fail this condition.  We should therefore not identify the eight algebraic sheets naively with eight real saddles of the original Hermitian integral.

Nevertheless, the extra algebraic sheets should not simply be discarded. Non-perturbative physics is very often controlled by saddles that become visible only after complexifying the original integration problem.  The familiar instanton is an example: it appears after analytic continuation to Euclidean signature, but it has genuine physical consequences.  In the same way, the algebraic sheets that do not lie on the original Hermitian contour remain natural candidate sectors of the complexified invariant theory.

We have therefore obtained the same structural picture from two rather different quantizations.  In the Hilbert-space description the primary invariants generate arbitrary perturbative excitations above the secondary states.  In the integral description the primary invariants are the continuous variables that are integrated over, while the secondary data label the discrete branches that must be summed over.  A perturbative expansion is local and explores fluctuations about one such branch.  Reconstructing the full invariant theory requires the complete discrete sum.

This is the main lesson of the zero-dimensional example.  The Hironaka decomposition is not merely a useful way of organizing polynomials.  It knows about a distinction which looks exactly like the familiar distinction between perturbative fluctuations and non-perturbative sectors.  In the next section we will ask how far this interpretation can be pushed when we return to quantum mechanics and genuine semiclassical sectors.

\subsection{Semiclassical Sectors}

In this section we again want to return to the connection between semiclassical sectors and secondary invariants, but now we want to do so at the level of quantum mechanics, which is a step up from our $0$-dimensional theory. When we talk about a semiclassical sector, we mean a region of configuration space organized around a classical saddle such that
\begin{itemize}
\item perturbation theory can be developed locally around that saddle;
\item the saddle is separated from other relevant saddles by an action or energy barrier;
\item transitions between the corresponding localized states are nonperturbative in the semiclassical parameter, typically of order $e^{-S_{\rm inst}/\hbar}$.
\end{itemize}
Clean examples of semiclassical sectors include distinct degenerate or nearly degenerate classical minima. They may be true vacua, metastable vacua or more general long-lived classical configurations. We can also speak of semiclassical sectors associated with nontrivial classical solutions, as long as the solution has no unstable directions.

We have already seen that fixing the values of the primary invariants leaves a finite set of distinct gauge-invariant configurations with identical primary data. To show that these fiber points behave as distinct semiclassical sectors, the natural dynamical test is:
\begin{itemize}
\item \emph{Classical realization:} Are the different physical fiber points realized as classical stationary configurations of the chosen action?
\item \emph{Stability:} Check (with the Hessian) whether each is a local minimum, or at least a sufficiently long-lived metastable saddle.
\item \emph{Degeneracy in primary observables:} Verify that the different configurations agree in their primary data but differ in one or more secondary invariants.
\item \emph{Barrier separation:} Establish that a continuous path between the configurations must cross a region of larger potential or Euclidean action.
\item \emph{Instanton interpolation:} Look for finite-action Euclidean solutions interpolating between configurations. Their actions determine exponentially small mixing between localized semiclassical states.
\end{itemize}
This would give a concrete meaning to the statement that secondary invariants are associated with semiclassical sectors. To perform this analysis, we need to choose the dynamics. We will study the quantum mechanics of the action
\begin{equation}
 S=\int d\tau\left[ \frac14\sum_{a=1}^{4}\Tr(\dot X^a\dot X^a)-V(X^a)\right]
\end{equation}
where we choose the positive quartic potential, which depends only on the primary invariants
\begin{eqnarray}
V(X^a)&=&{\rm Tr}\Bigg[
\frac{\lambda}{8}\left(X_1^2-\mathbf 1\right)^2+\frac{\lambda}{8}\left(X_3^2-\mathbf 1\right)^2+\frac{\lambda}{16}\left(\{X_2,X_4\}-\frac{1}{2}\left(X_2^2+X_4^2\right)\right)^2\nonumber\\[2mm]
&&\,\,\,\,+\frac{\lambda}{32}\left(X_2^2+X_4^2-4\mathbf 1\right)^2+\frac{\eta}{4}\left(X_2^2-X_4^2\right)^2
+\frac{\kappa}{16}\Big(\{X_1,X_2\}-\{X_3,X_4\}\Big)^2\nonumber\\[2mm]
&&\,\,\,\,+\frac{\rho}{16}\Big(\{X_1,X_3\}^2+\{X_1,X_4\}^2+\{X_2,X_3\}^2\Big)\Bigg]
\end{eqnarray}
Although this potential might look very complicated, it is chosen so that it fixes the values of the primary invariants and so that it does not depend on the secondary invariant which is given by $\Sigma=\vec{x}_1\cdot\vec{x}_2+\vec{x}_3\cdot\vec{x}_4$. The potential fixes the values of the primary invariants to
\begin{equation}
|\vec{x}_1|=|\vec{x}_3|=\vec{x}_2\cdot\vec{x}_4=1\qquad |\vec{x}_2|=|\vec{x}_4|=\sqrt{2}\qquad \vec{x}_1\cdot\vec{x}_3=\vec{x}_1\cdot\vec{x}_4=\vec{x}_2\cdot\vec{x}_3=0\nonumber
\end{equation}
Notice that the potential localizes us to the point whose physical fiber we analysed in detail in Section \ref{SecAsDisc}. Further, since the potential is independent of the secondary invariants, this minimum will appear on each sheet. Recall that the physical fiber is labelled by
\begin{equation}
(s,\epsilon)\in\{+1,-1\}\times\{+1,-1\}, \qquad \Sigma=s.\nonumber
\end{equation}
To switch between two roots of $\Sigma$ we need to switch between two values of $s$ at fixed $\epsilon$. A path that changes $s$ at fixed orientation is ($2\Sigma=\vec{x}_1\cdot\vec{x}_2+\vec{x}_3\cdot\vec{x}_4$, as long as $q(\tau)$ moves between $\pm 1$)
\begin{equation}
\vec x_1=q(\tau)\vec e_1,\quad \vec x_2=\vec e_1+\epsilon\vec e_3,\quad\vec x_3=q(\tau)\vec e_2,\quad
\vec x_4=\vec e_2+\epsilon\vec e_3.
\end{equation}
Using this as an ansatz for a solution to the Euclidean equations of motion, we find the following exact solution
\begin{equation}
q_{\rm Gram}(\tau)=\tanh\!\left[\sqrt{\lambda\over2}(\tau-\tau_0)\right],
\end{equation}
which has a positive and real action
\begin{equation}
S_{\rm Gram}={4\sqrt{2\lambda}\over3}.
\end{equation}
Since the action is real and positive, the mixing between the states with different roots for $\Sigma$ is highly suppressed by the factor $\sim e^{-S_{\rm Gram}/\hbar}$ and it is non-perturbative since this factor can not be expanded in a power series in $\hbar$. It is also possible to find a path that changes the orientation $\epsilon$ for a fixed root $\Sigma$. The path is given by
\begin{equation}
\vec x_a(\tau)=z(\tau)\vec x_a^{(s,+)},\qquad T_{abc}(\tau)=z(\tau)^3T_{abc}^{(s,+)},
\end{equation}
where again we assume that the unknown function $z(\tau)$ moves between $\pm 1$ and $\mp 1$. Using this as an ansatz for a solution to the Euclidean equations of motion, we easily find an exact solution
\begin{equation}
z_{\rm orient}(\tau)=\tanh\!\left[\sqrt{\lambda\over2}(\tau-\tau_0)\right]\,.
\end{equation}
The instanton again has a positive and real action
\begin{equation}
S_{\rm orient}=4\sqrt{2\lambda}.
\end{equation}
so that the mixing is highly suppressed $\sim e^{-S_{\rm orient}/\hbar}$ and non-perturbative.

This example demonstrates that the secondary labels can become genuine semiclassical sectors with exponentially suppressed mixing. Does this lesson carry over to more ``realistic'' settings? In particular, we would like to see if these lessons are also applicable at larger values of $N$ and for a generic potential. The potential above was engineered to: (i) fix the values of all of the primary invariants and (ii) to leave the values of the secondary invariants free. The property (ii) played a crucial role in our analysis since it ensured that the ground state is repeated in each secondary sheet. Further, it is easier to satisfy as $N$ is increased, since any degree bounded potential will then lie well below the overcrowding scale. The property (i) is needed to obtain a definite ground state. This will be satisfied for any models that have a set of discrete ground states. A good example of such a model is the BMN matrix model.

\section{Discussion}

Our goal in these lectures has been to understand the consequences of the trace relations, identities which capture genuine finite $N$ physics. We have explained that it is possible to impose the complete set of trace relations on the space of gauge invariant operators, to obtain a non-redundant but complete set of gauge invariant operators. The resulting set has a rich structure, that is nevertheless mathematically rather elegant. The complete set of gauge invariant operators is generated by primary and secondary invariants. Using these invariants to organize the Hilbert space of multi-matrix oscillators, it is evident that each primary invariant behaves as a creation operator, so the primary invariants generate a Fock space of states. There are $1+(d-1)N^2$ such primary invariants for a $d$ matrix model of $N\times N$ matrices. The secondary invariants generate states, each of which supports a primary invariant Fock space, so they are naturally identified with semi-classical sectors. The number of secondary invariants, which grows as $e^{cN^2}$ with $c$ an $O(1)$ number, is big enough to explain the entropy of a black hole. We have argued, in specific examples, that this interpretation is also natural within a path integral quantization of the theory.

There are many directions in which this work can be extended. We will mention below a few questions that already seem within reach.
\begin{itemize}
\item Bilocal algebras: An alternative approach to finite $N$ is to study the algebra of the invariant fields themselves. For the case of vector models, this leads to bilocal invariant variables for which it is possible to establish an invariant dual-pair operator algebra. The finite-$N$ trace relations become representation-theoretic identities selecting a particular irreducible representation. Finite traces and partition functions are realized as characters of the resulting irreducible representations~\cite{vatsal}.
\item Resurgence: The Hironaka decomposition summarizes global algebraic information about a finite-$N$ invariant ring, while resurgence reconstructs a global analytic answer from perturbative data defined locally around one saddle. How much of the global information assembled by resurgence is already encoded in the Hironaka decomposition?
\item Fortuity: Secondary invariants provide the invariant-theoretic, purely bosonic precursor of fortuity. Both arise because finite-$N$ trace relations change the structure of the physical operator space relative to its freely generated large-$N$ covering. Fortuitous states are the cohomological realization of this phenomenon when a supercharge is present.
\item Quantum Field Theory: In quantum field theory there are two highly nontrivial generalizations. First, trace relations may themselves be modified by renormalization, since composite operators must be defined in the presence of UV divergences and can mix under renormalization. Second, the finite-generation assumptions underlying the usual Hilbert–Serre and Hironaka framework are no longer automatic: there need not be a finite set of primaries and secondaries generating the full algebra of local operators.
\end{itemize}
We leave these fascinating questions for the future.

\section*{Acknowledgments}

This review is based on lectures delivered at the Second Joint Summer School on Theoretical High-Energy Physics (Beijing 2026) 8/16-28/2026, and on special lectures delivered at Kyung Hee University. We are very grateful to Ping Gao, Yunfeng Jiang and Xinan Zhou, for hospitality in Beijing, and to Euihun Joung and Junggi Yoon for hospitality in Seoul. Thanks also to the participants at the lectures for creating a stimulating scientific atmosphere. The work of RdMK and AR is supported by a start-up research fund of Huzhou Normal University, a Zhejiang Province talent award and a Changjiang Scholar award. The work of MK is supported by the NRF grants funded by the Korea government (MSIT) RS-2025-25414114.


\begin{thebibliography}{9}

\bibitem{AdSCFT}
J.~M.~Maldacena, ``The Large $N$ limit of superconformal field theories and supergravity,''
Adv. Theor. Math. Phys. \textbf{2} (1998), 231-252 [arXiv:hep-th/9711200 [hep-th]],\\
E.~Witten, ``Anti de Sitter space and holography,''
Adv. Theor. Math. Phys. \textbf{2} (1998), 253-291 [arXiv:hep-th/9802150 [hep-th]],\\
S.~S.~Gubser, I.~R.~Klebanov and A.~M.~Polyakov, ``Gauge theory correlators from noncritical string theory,''
Phys. Lett. B \textbf{428} (1998), 105-114 [arXiv:hep-th/9802109 [hep-th]].

\bibitem{SEP}
J.~M.~Maldacena and A.~Strominger, ``AdS(3) black holes and a stringy exclusion principle,'' JHEP \textbf{12} (1998), 005 [arXiv:hep-th/9804085 [hep-th]],\\
J.~McGreevy, L.~Susskind and N.~Toumbas, ``Invasion of the giant gravitons from Anti-de Sitter space,''
JHEP \textbf{06} (2000), 008 [arXiv:hep-th/0003075 [hep-th]].

\bibitem{GGE}
D.~Gaiotto and J.~H.~Lee, ``The giant graviton expansion,'' JHEP \textbf{08} (2024), 025 [arXiv:2109.02545 [hep-th]],\\
J.~H.~Lee, ``Exact stringy microstates from gauge theories,'' JHEP \textbf{11} (2022), 137 [arXiv:2204.09286 [hep-th]],\\
J.~H.~Lee, ``Trace relations and open string vacua,'' JHEP \textbf{02} (2024), 224
[arXiv:2312.00242 [hep-th]],\\
J.~H.~Lee and W.~Li, ``AdS$_3$ Quantum Gravity and Finite $N$ Chiral Primaries,''
[arXiv:2511.00636 [hep-th]],\\
P.~Caputa and R.~de Mello Koch, ``Gauge invariants at arbitrary N and trace relations,'' JHEP \textbf{12} (2025), 165 [arXiv:2509.05834 [hep-th]].

\bibitem{fort}
C.~M.~Chang and Y.~H.~Lin, ``Words to describe a black hole,'' JHEP \textbf{02} (2023), 109 [arXiv:2209.06728 [hep-th]].

\bibitem{evid}
S.~Choi, S.~Kim, E.~Lee and J.~Park, ``The shape of non-graviton operators for SU(2),'' JHEP \textbf{09} (2024), 029
[arXiv:2209.12696 [hep-th]],\\
S.~Choi, S.~Kim, E.~Lee, S.~Lee and J.~Park, ``Towards quantum black hole microstates,'' JHEP \textbf{11} (2023), 175
[arXiv:2304.10155 [hep-th]],\\
C.~M.~Chang, L.~Feng, Y.~H.~Lin and Y.~X.~Tao, ``Decoding stringy near-supersymmetric black holes,'' SciPost Phys. \textbf{16} (2024) no.4, 109 [arXiv:2306.04673 [hep-th]],\\
J.~Choi, S.~Choi, S.~Kim, J.~Lee and S.~Lee, ``Finite N black hole cohomologies,'' JHEP \textbf{12} (2024), 029
[arXiv:2312.16443 [hep-th]],\\
C.~M.~Chang and Y.~H.~Lin, ``Holographic covering and the fortuity of black holes,'' [arXiv:2402.10129 [hep-th]],\\
S.~Kim, J.~Lee, S.~Lee and H.~Oh, ``BPS phases and fortuity in higher spin holography,'' [arXiv:2511.03105 [hep-th]],\\
A.~Gadde, E.~Lee, R.~Raj and S.~Tomar, ``Probing non-graviton spectra in $ \mathcal{N}=4 $ SYM via BMN truncation and S-duality,'' JHEP \textbf{02} (2026), 026
[arXiv:2506.13887 [hep-th]],\\
C.~M.~Chang and Y.~H.~Lin, ``Violation of S-duality in classical $Q$-cohomology,''
[arXiv:2510.24008 [hep-th]],\\
J.~Choi and E.~Lee,``Konishi lifts a black hole,'' [arXiv:2511.09519 [hep-th]],\\
S.~Kim, S.~Kim, S.~Lee and J.~Park, ``Quantum black hole cohomologies,''
[arXiv:2606.27955 [hep-th]].

\bibitem{withantal}
R.~de Mello Koch and A.~Jevicki, ``Structure of loop space at finite N,'' JHEP \textbf{06} (2025), 011
[arXiv:2503.20097 [hep-th]]\\
R.~de Mello Koch and A.~Jevicki, ``Hilbert space of finite N multi-matrix models,'' JHEP \textbf{11} (2025), 145
[arXiv:2508.11986 [hep-th]].

\bibitem{animik}
R.~de Mello Koch, A.~Ghosh and H.~J.~R.~Van Zyl, ``Bosonic fortuity in vector models,'' JHEP \textbf{06} (2025), 246
[arXiv:2504.14181 [hep-th]].

\bibitem{minkyoo}
R.~de Mello Koch, M.~Kim and H.~J.~R.~Van Zyl, ``From symmetry to structure: gauge-invariant operators in multi-matrix quantum mechanics,'' JHEP \textbf{01} (2026), 031 [arXiv:2507.01219 [hep-th]].

\bibitem{anik}
R.~de Mello Koch, A.~Jevicki, G.~Kemp and A.~Rudra, ``Collective theory at finite-N: reduction of the emergent Hilbert space,'' JHEP \textbf{04} (2026), 048 [arXiv:2510.22525 [hep-th]].

\bibitem{withjoao}
R.~de Mello Koch and J.~P.~Rodrigues, ``Secondary invariants and non-perturbative states,'' JHEP \textbf{06} (2026), 251 [arXiv:2604.15600 [hep-th]].

\bibitem{withaniklarweh}
R.~de Mello Koch, A.~Rudra and A.~L.~Mahu, ``Overcrowding and the Finite-$N$ Hilbert Space,''
[arXiv:2607.23025 [hep-th]].

\bibitem{SMT}
D.~E.~Berenstein, J.~M.~Maldacena and H.~S.~Nastase, ``Strings in flat space and pp waves from N=4 superYang-Mills,'' JHEP \textbf{04} (2002), 013 [arXiv:hep-th/0202021 [hep-th]],\\
T.~Harmark and M.~Orselli, ``Spin Matrix Theory: A quantum mechanical model of the AdS/CFT correspondence,'' JHEP \textbf{11} (2014), 134 [arXiv:1409.4417 [hep-th]],\\
S.~Baiguera, T.~Harmark and Y.~Lei, ``The Panorama of Spin Matrix theory,''
JHEP \textbf{04} (2023), 075 [arXiv:2211.16519 [hep-th]].

\bibitem{sturmfels}
Bernd Sturmfels, “Algorithms in invariant theory,” Springer Science \& Business Media, (2008),\\
B. Grinstein, X. Lu, L. Merlo and P. Qu´ılez, “Hilbert series for covariants and their applications to minimal flavor violation,” JHEP 2024 (2024), 154 [arXiv:2312.13349 [hep-ph]],\\
Cox, D.A., O'shea, D. and Little, J., ``Ideals, varieties, and algorithms: an introduction to computational algebraic geometry and commutative algebra,'' (Vol. 10), 2007, New York: Springer.

\bibitem{procesi}
C. Procesi, ``The invariant theory of n × n matrices,'' Adv. Math. 19 (1976) 306–381,\\
Yu. P. Razmyslov, ``Trace identities of full matrix algebras over a field of characteristic zero,'' Math. USSR-Izv. 8 (1974) 727–760,\\
V. Drensky and E. Formanek, ``Polynomial Identity Rings,'' Advanced Courses in Mathematics CRM Barcelona, Birkh\"auser (2004).

\bibitem{4MW}
B.~Sundborg, ``The Hagedorn transition, deconfinement and N=4 SYM theory,''
Nucl. Phys. B \textbf{573} (2000), 349-363 [arXiv:hep-th/9908001 [hep-th]],\\
O.~Aharony, J.~Marsano, S.~Minwalla, K.~Papadodimas and M.~Van Raamsdonk,
``The Hagedorn - deconfinement phase transition in weakly coupled large N gauge theories,''
Adv. Theor. Math. Phys. \textbf{8} (2004), 603-696 [arXiv:hep-th/0310285 [hep-th]],\\
A.~T.~Kristensson and M.~Wilhelm, ``From Hagedorn to Lee-Yang: partition functions of $ \mathcal{N} $ = 4 SYM theory at finite N,'' JHEP \textbf{10} (2020), 006 [arXiv:2005.06480 [hep-th]],\\
J.~Pasukonis and S.~Ramgoolam, ``Quivers as Calculators: Counting, Correlators and Riemann Surfaces,''
JHEP \textbf{04} (2013), 094 [arXiv:1301.1980 [hep-th]].

\bibitem{TIS}
D.~A.~McGady, ``Temperature-reflection I: field theory, ensembles, and interactions,'' [arXiv:1711.07536 [hep-th]],\\
D.~A.~McGady, ``Temperature-reflection II: Modular Invariance and T-reflection,''
[arXiv:1806.09873 [hep-th]].

\bibitem{HochesterRoberts}
M. Hochster and J. L. Roberts, ``Rings of invariants of reductive groups acting on regular rings are Cohen–Macaulay,'' Adv. Math. 13 (1974) 115–175.

\bibitem{paledrmic}
Y. Teranishi, ``The Hilbert series of rings of matrix concomitants'', Nagoya Mathematical Journal, 111, (1988), pp.143-156,\\
Y. Teranishi, ``The ring of invariants of matrices,'' Nagoya Mathematical Journal, 104, (1986), pp.149-161,\\
Y. Teranishi, ``Linear Diophantine equations and invariant theory of matrices,'' In Commutative algebra and combinatorics (Vol. 11, 1987, pp. 259-276). Mathematical Society of Japan.\\
D.Z. Dokovi\'c, "Poincaré series of some pure and mixed trace algebras of two generic matrices." Journal of Algebra 309, no. 2 (2007): 654-671.

\bibitem{RSP}
R.~de Mello Koch, J.~Smolic and M.~Smolic, ``Giant Gravitons - with Strings Attached (I),''
JHEP \textbf{06} (2007), 074 [arXiv:hep-th/0701066 [hep-th]],\\
R.~Bhattacharyya, S.~Collins and R.~de Mello Koch, ``Exact Multi-Matrix Correlators,''
JHEP \textbf{03} (2008), 044 [arXiv:0801.2061 [hep-th]],\\
R.~Bhattacharyya, R.~de Mello Koch and M.~Stephanou, ``Exact Multi-Restricted Schur Polynomial Correlators,''
JHEP \textbf{06} (2008), 101 [arXiv:0805.3025 [hep-th]],\\
R.~de Mello Koch, P.~Diaz and N.~Nokwara, ``Restricted Schur Polynomials for Fermions and integrability in the $su(2|3)$ sector,'' JHEP \textbf{03} (2013), 173 [arXiv:1212.5935 [hep-th]].

\bibitem{Sanjaye}
Y.~Kimura and S.~Ramgoolam, ``Branes, anti-branes and brauer algebras in gauge-gravity duality,''
JHEP \textbf{11} (2007), 078 [arXiv:0709.2158 [hep-th]],\\
T.~W.~Brown, P.~J.~Heslop and S.~Ramgoolam, ``Diagonal multi-matrix correlators and BPS operators in N=4 SYM,''
JHEP \textbf{02} (2008), 030 [arXiv:0711.0176 [hep-th]],\\
T.~W.~Brown, P.~J.~Heslop and S.~Ramgoolam, ``Diagonal free field matrix correlators, global symmetries and giant gravitons,'' JHEP \textbf{04} (2009), 089 [arXiv:0806.1911 [hep-th]],\\
S.~Ramgoolam, ``Schur-Weyl duality as an instrument of Gauge-String duality,'' AIP Conf. Proc. \textbf{1031} (2008) no.1, 255-265 [arXiv:0804.2764 [hep-th]],\\
J.~Pasukonis and S.~Ramgoolam, ``From counting to construction of BPS states in N=4 SYM,''
JHEP \textbf{02} (2011), 078 [arXiv:1010.1683 [hep-th]],\\
Y.~Kimura, ``Correlation functions and representation bases in free N=4 Super Yang-Mills,''
Nucl. Phys. B \textbf{865} (2012), 568-594 [arXiv:1206.4844 [hep-th]].

\bibitem{RSPintro}
R.~de Mello Koch, M.~Kim and A.~L.~Mahu, ``A pedagogical introduction to restricted Schur polynomials with applications to heavy operators,'' Int. J. Mod. Phys. A \textbf{39} (2024) no.31, 2430003 [arXiv:2409.15751 [hep-th]].

\bibitem{PakPanova}
I. Pak, G. Panova and D. Yeliussizov, “On the largest Kronecker and Littlewood–Richardson
coefficients,” Journal of Combinatorial Theory, Series A 165 (2019): 44-77.

\bibitem{VKLS}
A. M. Vershik and S. V. Kerov, ``Asymptotic of the largest and the typical dimensions of irreducible representations of a symmetric group,'' Funct. Anal. Appl. 19 (1985), 21–31,\\
B. F. Logan and L. A. Shepp, ``A variational problem for random Young tableaux,'' Adv. Math. 26 (1977), 206–222.

\bibitem{collective}
A.~Jevicki and B.~Sakita, ``The Quantum Collective Field Method and Its Application to the Planar Limit,''
Nucl. Phys. B \textbf{165} (1980), 511,\\
A.~Jevicki and B.~Sakita, ``Collective Field Approach to the Large $N$ Limit: Euclidean Field Theories,''
Nucl. Phys. B \textbf{185} (1981), 89-100.

\bibitem{constructiveholog}
S.~R.~Das and A.~Jevicki, ``String Field Theory and Physical Interpretation of $D=1$ Strings,''
Mod. Phys. Lett. A \textbf{5} (1990), 1639-1650,\\
S.~R.~Das and A.~Jevicki, ``Large N collective fields and holography,''
Phys. Rev. D \textbf{68} (2003), 044011 [arXiv:hep-th/0304093 [hep-th]],\\
R.~de Mello Koch, A.~Jevicki, K.~Jin and J.~P.~Rodrigues, ``$AdS_4/CFT_3$ Construction from Collective Fields,''
Phys. Rev. D \textbf{83} (2011), 025006 [arXiv:1008.0633 [hep-th]],\\
R.~de Mello Koch, ``Gravitational dynamics from collective field theory,''
JHEP \textbf{10} (2023), 151 [arXiv:2309.11116 [hep-th]],\\
R.~de Mello Koch and H.~J.~R.~Van Zyl, ``Constructive holography,'' JHEP \textbf{09} (2024), 022
[arXiv:2406.18248 [hep-th]].

\bibitem{failfact}
D.~Garner, S.~Ramgoolam and C.~Wen, ``Thresholds of large N factorization in CFT$_{4}$: exploring bulk spacetime in AdS$_{5}$,'' JHEP \textbf{11} (2014), 076 [arXiv:1403.5281 [hep-th]].

\bibitem{fastscramblers}
Y.~Sekino and L.~Susskind, ``Fast Scramblers,'' JHEP \textbf{10} (2008), 065 [arXiv:0808.2096 [hep-th]].

\bibitem{ballistic}
E. H. Lieb and D. W. Robinson, “The finite group velocity of quantum spin systems,” Commun. Math. Phys. 28 (1972), 251-257,\\
M. B. Hastings and T. Koma, “Spectral gap and exponential decay of correlations,” Commun. Math. Phys. 265 (2006), 781-804 [arXiv:math-ph/0507008 [math-ph]],\\
M. B. Hastings, “Locality in Quantum Systems,” [arXiv:1008.5137 [math-ph]].

\bibitem{colemansemiclassical}
For a classical reference on the semiclassical expansion in path integrals see Chapter 7 of S. Coleman, ``Aspects of Symmetry.'' 

\bibitem{vatsal}
R.~de Mello Koch, A.~Jevicki and J.~Yoon, ``Finite N Hilbert spaces of bilocal holography,''
JHEP \textbf{07} (2026), 084 [arXiv:2602.20788 [hep-th]],\\
B.~Ahn, R.~de Mello Koch, V.~Garg, A.~Jevicki and J.~Yoon, ``Finite-$N$ Operator Algebras and the Hilbert Space of Bilocal Holography,'' [arXiv:2607.04143 [hep-th]].
\end{thebibliography}
\end{document}